\documentclass[acmsmall,10pt,screen]{acmart}
\usepackage{ifpdf}
\ifpdf
\pdfoutput=1 
\fi
\usepackage{booktabs} 
\usepackage{pgf-umlsd}

\usepackage[ruled]{algorithm2e} 

\usepackage{cleveref}
\crefname{lstlisting}{listing}{listings}
\Crefname{lstlisting}{Listing}{Listings}

\usepackage{listings}

\lstdefinestyle{Python3}{
  morekeywords={yield,await,async}
}
\usepackage{color}
\usepackage{dot2texi}
\usepackage{tikz}
\usetikzlibrary{shapes,arrows}

\usepackage[inline]{enumitem}

\usepackage{tabulary}

\setcopyright{none}

\begin{document}
\title{Ad-hoc polymorphic delimited continuations}
\subtitle{unifying monads and continuations}

\author{Yang, Bo}
\orcid{0000-0003-2757-9115}
\affiliation{%
  \institution{ThoughtWorks, Inc}
}
\email{atryyang@thoughtworks.com}

\begin{abstract}
We designed and implemented a framework for creating extensible domain-specific languages that consists of library-defined keywords.
First-class language features in other programming languages can be implemented as libraries with the help of our framework.

The core concept in our framework is the type class \texttt{Dsl}, which can be considered as both the ad-hoc polymorphic version of a delimited continuation and a more generic version of \texttt{Monad}. Thus it can be also used as a statically typed extensible effect system that is more efficient and more concise than existing \texttt{Monad}-based effect systems.
\end{abstract}

%
%


\maketitle

\section{Introduction}\label{Introduction}

Traditionally, the capacity of a general purpose language can be extended to a special domain by creating an embedded Domain-Specific Language (eDSL) \cite{fowler2010domain} . For example, Akka provides a DSL to create finite-state machines \cite{lightbend2017akka}, which consists of some domain-specific operators including \lstinline{when}, \lstinline{goto}, \lstinline{stay}, etc. Although those operators looks similar to native control flow, they are not embeddable in native \lstinline{if}, \lstinline{while} or \lstinline{try} blocks, because the DSL code is split into small closures, preventing ordinary control flow from crossing the boundary of those closures. Thus, this kind of DSLs reinvent incompatible control flow to the meta-languages. TensorFlow's control flow operations \cite{abadi2016tensorflow} and Caolan's async library \cite{caolan2017async} are other examples of reinventing control flow in eDSLs.

Instead of reinventing the whole set of control flow for each DSL, a more general approach is designing a common protocol for control flow operators of all domains. In Haskell, Scala, and other functional programming language, monads are used as the generic protocol of control flow operators \cite{wadler1990comprehending,wadler1992essence,jones1993composing}. Scala implementations of monads are provided by Scalaz \cite{kenji2017scalaz}, Cats \cite{typelevel2017cats}, Monix \cite{nedelcu2017monix} and Algebird \cite{twitter2016algebird}.  A DSL author only has to implement \lstinline{>>=} and \lstinline[language=Haskell,deletekeywords={return}]{return} operators in \lstinline{Monad} type class, and all the derived control flow operations like \lstinline{whileM} or \lstinline{ifM} are available. In addition, those monadic data types can be created and composed from \lstinline{do}-notation \cite{jones1998haskell} or \lstinline{for}-comprehension \cite{odersky2004scala}. For example, in Scala, you can use the same \lstinline{scalaz.syntax} or \lstinline{for}-comprehension to create random value generators \cite{nilsson2015scalacheck} and data-binding expressions \cite{yangbo2016binding}, as long as there are \lstinline{Monad} instances for those domain-specific monadic data types respectively.

An idea to avoid incompatible domain-specific control flow is converting direct style control flow to domain-specific control flow at compile time. For example, Scala Async provides a macro to generate asynchronous control flow \cite{haller2013sip}, allowing normal sequential code inside a \lstinline{scala.async} block to run asynchronously. This approach can be generalized to any monadic data types. ThoughtWorks Each \cite{yangbo2015each}, Monadless \cite{flavio2017monadless}, effectful \cite{crockett2013effectful} and !-notation in Idris \cite{brady2013idris} are compiler-time transformers to convert source code of direct style control flow to monadic control flow. For example, with the help of ThoughtWorks Each, Binding.scala \cite{yangbo2016binding} can be used to create reactive HTML template from ordinary direct style code.

Another generic protocol of control flow is delimited continuation, which is known as the mother of all monads \cite{filinski1994representing,piponi2008mother}, where specific control flow in specific domain can be supported by specific answer types of continuations \cite{asai2007polymorphic}. Scala Continuations \cite{rompf2009implementing} and Stateless Future \cite{yangbo2014stateless} are two delimited continuation implementations in Scala. Both projects can convert direct style control flow to continuation-passing style closure chains at compile time. For example, Stateless Future Akka \cite{yangbo2014statelessfutureakka}, based on Stateless Future, provides a special answer type for akka actors. Unlike reinvented control flow in \lstinline{akka.actor.AbstractFSM}, users can create complex finite-state machines from simple direct style control flow along with Stateless Future Akka's domain-specific operator \lstinline{nextMessage}.

All the previous approaches lack of the ability to collaborate with other DSLs. Each of the above DSLs can be exclusively enabled in a code block. Scala Continuations enables calls to \lstinline{@cps} method in \lstinline{reset} blocks, and ThoughtWorks Each enables the magic \lstinline{each} method \cite{yangbo2015each} for \lstinline{scalaz.Monad} in \lstinline{monadic} blocks. It was impossible to enable both DSL in one function.

Monad transformers \cite{liang1995monad} is a popular technique to solve the collaboration problem. The basic idea is to use an ad-hoc polymorphic \lstinline{lift} function to convert different monadic type into the same transformed monadic type. Thus a \lstinline{do} block of a transformed monadic type can contain different DSL operations as long as they can be \lstinline{lift}ed. With the help of additional type classes, those \lstinline{lift} operations can be performed automatically.

However, a deeply nested transformed monad was considered inefficient  due to the nested \lstinline{lift}. An alternative approach proposed by
\cite{kiselyov2013extensible} is effect handlers. In the effect handler approach, the DSL ``script'' is written in a universal monadic type \lstinline{Eff}, which allows for multiple DSLs in one \lstinline{do} block. Each DSL is considered as an effect, which is dispatched by \lstinline{Eff} to the specific \lstinline{Handler}. This approach is heavy-weight, since only expressions written in \lstinline{Eff} script are able to use DSLs defined in effect handlers. Additional conversion is required to retrieve the ``raw'' data type from an \lstinline{Eff} \lstinline{do} block.

This paper proposes a new type class \lstinline{Dsl}, which can be considered as both the ad-hoc polymorphic version of a delimited-continuation and a more generic version of \lstinline{Monad}. The Scala definition of the type class is shown in \cref{Dsl}.

\begin{lstlisting}[caption={The definition of \lstinline{Dsl} type class},label={Dsl}]
trait Dsl[Keyword, Domain, Value] {
  def cpsApply(keyword: Keyword, handler: Value => Domain): Domain
}
\end{lstlisting}

\begin{figure}[h t b p]
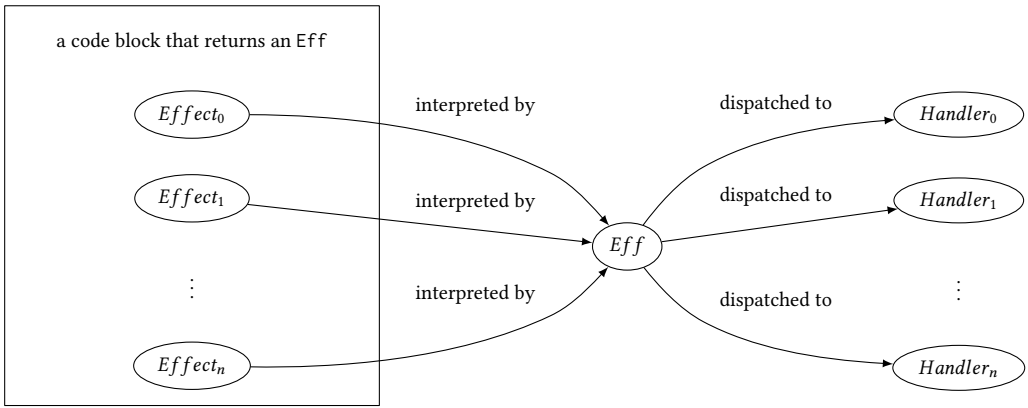

  \begin{dot2tex}[dot,mathmode,autosize,graphstyle={scale=0.78,transform shape}]
  digraph {
  	rankdir=LR
    shape=rect

  	subgraph cluster_code_block {
      graph [ label="\textrm{a code block that returns an \lstinline{Eff}}" ]
      
      Effect_0
      Effect_1
      Effect_dots [ label="{\vdots}" shape=none ]
      Effect_n
    }
    
    Effect_0 -> "Eff"  [ label="\textrm{interpreted by}" ]
    "Eff" -> Handler_0 [ label="\textrm{dispatched to}" ]
    Effect_1 -> "Eff"  [ label="\textrm{interpreted by}" ]
    "Eff" -> Handler_1 [ label="\textrm{dispatched to}" ]
    Handler_dots  [ label="{\vdots}" shape=none ]
    Effect_dots -> "Eff"  [ style=invis ]
    "Eff" -> Handler_dots [ style=invis ]
    Effect_n -> "Eff"  [ label="\textrm{interpreted by}" ]
    "Eff" -> Handler_n [ label="\textrm{dispatched to}" ]
  }
  \end{dot2tex}

  \caption{The architecture of \lstinline{Eff} approach}
  \label{eff-architecture}
\end{figure}

\begin{figure}[h t b p]
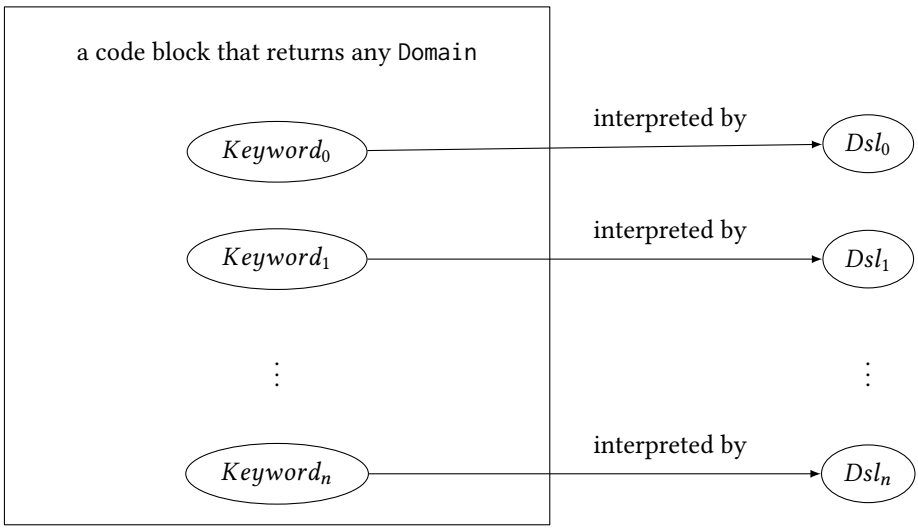

  \begin{dot2tex}[dot,mathmode,autosize,graphstyle={transform shape}]
  digraph {
  	rankdir=LR
    shape=rect

  	subgraph cluster_code_block {
      graph [ label="\textrm{a code block that returns any \texttt{Domain}}" ]
      Keyword_0
      Keyword_1
      Keyword_dots [ label="{\vdots}" shape=none ]
      Keyword_n
    }

    Keyword_0 -> Dsl_0 [ label="\textrm{interpreted by}" ]
    Keyword_1 -> Dsl_1 [ label="\textrm{interpreted by}" ]
    Keyword_dots -> Dsl_dots [ style=invis ]
    Keyword_n -> Dsl_n [ label="\textrm{interpreted by}" ]

    Dsl_dots [ label="{\vdots}" shape=none ]
  }
  \end{dot2tex}

  \caption{The architecture of \lstinline{Dsl} approach}
  \label{dsl-architecture}
\end{figure}

Because \lstinline{Dsl} is more generic than \lstinline{Monad}, it allows a code block to contain interleaved heterogeneous \lstinline{Keyword}s, interpreted by different \lstinline{Dsl} type class instances. Instead of returning an intermediate script type like \lstinline{Eff} \cite{kiselyov2013extensible}, the return types of a DSL code block are the final result type, which can vary as long as where are corresponding \lstinline{Dsl} instances for all operators inside the DSL code block. No intermediate \lstinline{Monad} for dispatching is used. The difference of architecture between effect handler approach and our approach is shown in \cref{eff-architecture,dsl-architecture}.

Our approach is more flexible than ordinary delimited continuation, too. An ordinary delimited continuation \cite{danvy1989functional} can be defined as a CPS (Continuation-Passing Style) functions to register a callback function (\cref{Continuation}), which is similar to the signature of \lstinline{Dsl} type class.

\begin{lstlisting}[caption={The definition of a delimited continuation},label={Continuation}]
type Continuation[Domain, Value] = (Value => Domain) => Domain
\end{lstlisting}

Since a \lstinline{Continuation} is a function, it contains the hard-coded implementation of an operation. As a result, a delimited continuation can only be used in a function that returns the specified \lstinline{Domain}. In contrast, in our approach, each \lstinline{Keyword} is ad-hoc polymorphic to the \lstinline{Domain}, thus it can be interpreted differently according to the enclosing \lstinline{Domain}.

In the remaining sections of this paper, we will present the design and use cases of \lstinline{Dsl} type class, including:

\begin{enumerate}
  \item Simulating some first-class features in Python, C\#, ECMAScript and C++, as library-defined keywords;
  \label{simulating-first-class-features}

  \item Simulating \lstinline{Monad} to create imperative code blocks;
  \label{simulating-monad}

  \item Composing delimited continuations with less closure creation than \lstinline{Monad} for continuations;
  \label{composing-delimited-continuations}

  \item Making \lstinline{Continuation} stack safe, in a non-intrusive way; 
  \label{stack-safe-Continuation}

  \item Using any combination of the features of \cref{simulating-first-class-features,simulating-monad,composing-delimited-continuations,stack-safe-Continuation}, in a single code block.
\end{enumerate}

All code examples except \cref{Haskell implementation} are written in our Scala library \textit{Dsl.scala}, which provides some built-in instances of \lstinline{Dsl} type class, along with a Scala compiler plug-in to perform a CPS-transformation. The compiler plug-in avoids the ``callback hell'' problem, allowing Idris-like !-notation \cite{brady2013idris} direct style DSL in Scala, which can be used for not only monadic data type but also other operations.

\section{From delimited continuation to the \lstinline{Dsl} type class}\label{Basic concepts}

Our goal is making the control flow of a programming language to be extensible. In this section, we will introduce the \lstinline{Dsl} type class and the concept of name-based CPS transformation. We will also demonstrate how to use these techniques to port first class Python language features to Scala, as library-defined keywords (LDK) \footnote{Code listings shown in \cref{Basic concepts} are not exactly the same as the implementation in \textit{Dsl.scala}, instead, these implementations of LDKs are modified or simplified for the purpose of introducing the concept of the LDK approach more clearly.}. The term LDK denotes language features implemented by libraries. No metaprogramming knowledge is required for either LDK authors or LDK users \footnote{ Though, Scala LDKs need the common compiler plug-ins to perform CPS transformation and Haskell LDKs need \lstinline{RebindableSyntax} described in \cref{Haskell implementation}}, while, in other languages, they are used to be implemented as compiler built-in first-class features.

The remaining parts of this section are organized as following. Firstly, in \cref{Implementing LDKs as ordinary delimited continuations,CPS transformation}, we will present how to port \lstinline{yield} to Scala in the ordinary delimited continuation approach. Then in \cref{Monadic blocks}, we will present how to port \lstinline{await} to Scala in a monad-like interface. Finally, in \cref{Collaborative library-defined keywords,Adaptive library-defined keywords}, we will introduce the type class \lstinline{Dsl} to unifying all the previous approaches, and in addition, allowing for the use of multiple LDKs like \lstinline{yield} and \lstinline{await} together.

\subsection{Implementing LDKs as ordinary delimited continuations}\label{Implementing LDKs as ordinary delimited continuations}

In Python, ECMAScript, and C\#, a generator is a function that returns an \lstinline{Iterator} or an \lstinline{IEnumerator}. The \lstinline{yield} keyword is available inside the generator to lazily produce one element, which can be consumed by the \lstinline{Iterator} / \lstinline{IEnumerator} user. \Cref{xorShiftRandomGenerator-Python} is a Python example to create an xorshift \cite{marsaglia2003xorshift} pseudo-random number generator that returns an infinite iterator of generated numbers. Note that NumPy \footnote{http://www.numpy.org/} is used for 32-bit integers, and type hinting \footnote{https://docs.python.org/3/library/typing.html} is used for clarity.

\begin{lstlisting}[language=Python,style=Python3,caption={An Xorshift pseudo-random number generator in Python 3.5+},label={xorShiftRandomGenerator-Python}]
def xor_shift_random_generator(seed: np.uint32) -> Iterator[np.uint32]:
  tmp1 = np.uint32(seed ^ (seed << 13))
  tmp2 = np.uint32(tmp1 ^ (tmp1 >> 17))
  tmp3 = np.uint32(tmp2 ^ (tmp2 << 5))
  yield tmp3
  yield from xor_shift_random_generator(tmp3)

generated_numbers = xor_shift_random_generator(seed = np.uint32(2463534242))

print(generated_numbers.__next__()) // The first generated random number
print(generated_numbers.__next__()) // The second generated random number
\end{lstlisting}

This generator feature can be ported to Scala as an LDK. In our LDK-based generator, the return type is replaced to \lstinline{scala.Stream}, which can be considered as the immutable version of \lstinline{Iterator}, and the compiler-defined keyword \lstinline{yield} is replaced to library-defined keyword \lstinline{Yield}. \Cref{xorShiftRandomGenerator} is an example to create an Xorshift \cite{marsaglia2003xorshift} pseudo-random number generator that returns an infinite stream of generated numbers. 

\lstinline{xorShiftRandomGenerator} does not throw a \lstinline{StackOverflowError}, because the execution of \lstinline{xorShiftRandomGenerator} will be paused at \lstinline{Yield}, and it will be resumed when the caller is looking for the next number.

\begin{lstlisting}[caption={An Xorshift pseudo-random number generator with the help of the LDK \lstinline{Yield}},label={xorShiftRandomGenerator}]
def xorShiftRandomGenerator(seed: Int): Stream[Int] = {
  val tmp1 = seed ^ (seed << 13)
  val tmp2 = tmp1 ^ (tmp1 >>> 17)
  val tmp3 = tmp2 ^ (tmp2 << 5)
  Yield(tmp3) { _: Unit =>
    xorShiftRandomGenerator(tmp3)
  }
}

val generatedNumbers = xorShiftRandomGenerator(seed = 2463534242)

println(generatedNumbers(0)) // The first generated random number
println(generatedNumbers(1)) // The second generated random number
\end{lstlisting}

Despite of the implementation of \lstinline{Yield}, which will be revealed in upcoming sections, the above use case demonstrates some basic concepts in our approach:

\begin{enumerate}
  \item \lstinline{xorShiftRandomGenerator}, and any other functions that contain nested continuation-passing style (CPS) calls, are considered as written in some kind of eDSL.

  \item The word ``domain'' in the term ``Domain-Specific Language'' stands for the return type of the enclosing function. For example, \lstinline{Stream[Int]} is the domain of \lstinline{xorShiftRandomGenerator}.
  \label{domain-definition}

  \item The domain-specific language used by the enclosing function consists of some domain-specific ``library-defined keywords'' (LDK). For example, \lstinline{Yield} is an LDK available for \lstinline{Stream} domains.
  \label{ldk-definition}

  \item Along with LDK, DSLs written in \textit{Dsl.scala} also support native Scala control flows and expressions.
\end{enumerate}

For a simple use case such as \lstinline{xorShiftRandomGenerator}, LDKs can be implemented as ordinary delimited continuations. \Cref{NonadaptiveYield} shows an implementation of the \lstinline{Yield} LDK, as a delimited continuation, in which the \lstinline{Yield} LDK creates infinite \lstinline{Stream}s by capturing \lstinline{handler} into a lazily evaluated \lstinline{Stream.Cons}.

\begin{lstlisting}[caption={Implementing \lstinline{Yield} LDK as an ordinary delimited continuation},label={NonadaptiveYield}]
case class Yield[A](element: A) extends Continuation[Stream[A], Unit] {
  def apply(handler: Unit => Stream[A]): Stream[A] = {
    new Stream.Cons(element, handler(()))
  }
}
\end{lstlisting}

\subsection{Auto-\lstinline{reset} name-based CPS transformation}\label{CPS transformation}

The syntax of \cref{xorShiftRandomGenerator} differs from first-class generators in Python, as the code block contains some manually created CPS closures. Ideally, the ``rest'' program after a \lstinline{Yield} operation should be indented at the same level of \lstinline{Yield}, not in a nested closure. This coding style can be achieved by the !-notation provided by \textit{Dsl.scala}'s built-in compiler plug-ins. The function \lstinline{xorShiftRandomGenerator} can be written as \cref{xorShiftRandomGenerator-bang} with the help of the !-notation plug-ins.

\begin{lstlisting}[caption={TheXorshift pseudo-random number generator,  in the style of !-notation},label={xorShiftRandomGenerator-bang}]
def xorShiftRandomGenerator(seed: Int): Stream[Int] = {
  val tmp1 = seed ^ (seed << 13)
  val tmp2 = tmp1 ^ (tmp1 >>> 17)
  val tmp3 = tmp2 ^ (tmp2 << 5)
  !Yield(tmp3)
  xorShiftRandomGenerator(tmp3)
}
\end{lstlisting}

Our compiler plug-ins performs CPS-transform in a similar approach to \lstinline{reset}/\lstinline{shift} control operators in Scala Continuations \cite{rompf2009implementing}. Our domain type corresponds to the answer type in delimited continuations; our \lstinline{!} prefix corresponds to the \lstinline{shift} control operator; and the \lstinline{reset} control operator will be automatically injected to every function body. Thus the above \lstinline{xorShiftRandomGenerator} is equivalent to \cref{xorShiftRandomGenerator-scala-continuations} in Scala Continuations.

\begin{lstlisting}[caption={TheXorshift pseudo-random number generator,  in Scala Continuations},label={xorShiftRandomGenerator-scala-continuations}]
def xorShiftRandomGenerator(seed: Int): Stream[Int] = reset {
  val tmp1 = seed ^ (seed << 13)
  val tmp2 = tmp1 ^ (tmp1 >>> 17)
  val tmp3 = tmp2 ^ (tmp2 << 5)
  shift(Yield(tmp3))
  xorShiftRandomGenerator(tmp3)
}
\end{lstlisting}

Because of the automatically injected \lstinline{reset} control operator, the boundary of a delimited continuation can never be escaped from a function in our approach. Therefore, our plug-ins are able to eliminate the internal context of delimited continuations. Scala Continuations' \lstinline{ControlContext} and \lstinline{cps} type annotations are not necessary any more.

There is another difference between our compiler plug-ins and Scala Continuation. Our compiler plug-ins are name-based instead of type-based, allowing CPS-transformation in monadic blocks, which will be discussed in next section.

\subsection{Monadic blocks}\label{Monadic blocks}

In previous sections, we have demonstrated how to port the compiler-defined keyword \lstinline{yield} to Scala, as a library-defined keyword. In this section, we will demonstrate how to import another compiler-defined keyword, \lstinline{await}, to Scala, as a library-defined keyword.

\lstinline{await} is available in Python, ECMAScript, or C\#, to compose multiple asynchronous tasks into one task. The compiler-defined keyword \lstinline{await} in Python is available in functions marked as \lstinline{async}. Each \lstinline{await} pauses the execution until the awaiting operation is completed, and each \lstinline{return} keyword in an \lstinline{async} function will turn the return value into an \lstinline{Awaitable}. An example of creating an \lstinline{Awaitable} to download two web pages by aiohttp \footnote{https://docs.aiohttp.org/} is shown in \cref{download_two_pages-Python}.

\begin{lstlisting}[language=Python,style=Python3,caption={Asynchronously downloading two web pages in Python},label={download_two_pages-Python}]
async def download_two_pages() -> Awaitable[Tuple[bytes, bytes]]:
  session = aiohttp.ClientSession()
  response1 = await session.get('http://example.com')
  content1 = await response1.read()
  response2 = await session.get('http://example.net')
  content2 = await response2.read()
  return (content1, content2)
\end{lstlisting}

When porting \lstinline{await} feature to Scala, we replaced the compiler-defined keyword \lstinline{await} to a library-defined keyword \lstinline{Await}, and replaced \lstinline{Awaitable} to \lstinline{Future} \footnote{https://docs.scala-lang.org/overviews/core/futures.html} as shown in \cref{downloadTwoPages}. Note that \lstinline{ByteString}, \lstinline{Http}, \lstinline{HttpMethods}, \lstinline{HttpRequest} in \lstinline{downloadTwoPages} are asynchronous HTTP library provided by Akka \footnote{https://akka.io/} and Akka HTTP \footnote{https://akka.io/akka-http/}.

\begin{lstlisting}[caption={Asynchronously downloading two web pages in \textit{Dsl.scala} },label={downloadTwoPages}]
def downloadTwoPages(): Future[(ByteString, ByteString)] = {
  Await(Http().singleRequest(HttpRequest(HttpMethods.GET, "http://example.com"))) { response1 =>
    Await(response1.entity.toStrict(timeout = 5.seconds)) { content1 =>
      Await(Http().singleRequest(HttpRequest(HttpMethods.GET, "http://example.net"))) { response2 =>
        Await(response2.entity.toStrict(timeout = 5.seconds)) { content2 =>
          Future((content1.data, content2.data))
        }
      }
    }
  }
}
\end{lstlisting}

\lstinline{Await} should accept a \lstinline{handler} to handle the incoming value in an asynchronous \lstinline{Future}, and it can be implemented as a forwarder of \lstinline{flatMap} on \lstinline{Future}, as shown in \cref{NonadaptiveAwait}.

\begin{lstlisting}[caption={Implementing \lstinline{Await} LDK as a forwarder to \lstinline{flatMap}},label={NonadaptiveAwait}]
case class Await[A](future: Future[A]) {
  def apply[B](handler: A => Future[B])(implicit ec: ExecutionContext): Future[B] = {
      future.flatMap(handler)
  }
}
\end{lstlisting}

Similar to CPS-transformation in \cref{xorShiftRandomGenerator-bang}, the nested callback functions registered to \lstinline{Await} in the \lstinline{downloadTwoPages} method can be replaced to !-notation with the help of our compiler plug-ins. The direct style version of \lstinline{downloadTwoPages} is shown in \cref{downloadTwoPages-bang}.

\begin{lstlisting}[caption={Asynchronously downloading two web pages, in the style of !-notation},label={downloadTwoPages-bang}]
def downloadTwoPages(): Future[(ByteString, ByteString)] = Future {
  val response1 = !Await(Http().singleRequest(HttpRequest(HttpMethods.GET, "http://example.com"))) 
  val content1 = !Await(response1.entity.toStrict(timeout = 5.seconds))
  val response2 = !Await(Http().singleRequest(HttpRequest(HttpMethods.GET, "http://example.net")))
  val content2 = !Await(response2.entity.toStrict(timeout = 5.seconds))
  (content1.data, content2.data)
}
\end{lstlisting}

Note that \cref{downloadTwoPages-bang} are unable to be expressed in Scala Continuation because the \lstinline{shift} control operator accepts only CPS-functions, while the signature of \lstinline{flatMap} differs from CPS-functions, due to the additional type parameter \lstinline{B} and the additional implicit parameter of \lstinline{ExecutionContext}.

Fortunately our CPS-transformation compiler plug-ins are name-based. Given any expression $e_0$, $e_1$, ..., $e_n$, variable name $v_0$, $v_1$, ..., $v_n$ and the final expression $r$ in a function \lstinline{f}, as shown in \cref{bang-block}, our compiler plug-ins will convert the code block to \cref{cps-block}. The plug-ins convert ! prefixes to callback functions registrations, regardless what the signatures of those expressions are. Both delimited continuation and monad-like operations are supported. The behavior of our CPS-transformation compiler plug-ins is similar to !-notation in Idris or \lstinline{do}-notation with \lstinline{RebindableSyntax} in Haskell.

\begin{lstlisting}[mathescape=true, caption={A function with !-notation}, label={bang-block}]
def f = {
  val $v_0$ = !$e_0$;
  val $v_1$ = !$e_1$;
  $\hdots$
  val $v_n$ = !$e_n$;
  $r$;
}
\end{lstlisting}

\begin{lstlisting}[mathescape=true,caption={The code converted from !-notation by our name-based CPS-transformation plug-ins}, label={cps-block}]
def f = {
  $e_0$ { $v_0$ =>
    $e_1$ { $v_1$ =>
      $\hdots$
      $e_n$ { $v_n$ =>
        $r$
      }
    }
  }
}
\end{lstlisting}

While \lstinline{Await} implemented in \cref{NonadaptiveAwait} can ``extract'' the value of a \lstinline{Future}, it can be generalized to any \lstinline{Monad}s as shown in \cref{NonadaptiveMonadic}.

\begin{lstlisting}[caption={Implementing \lstinline{Monadic} LDK as a forwarder to \lstinline{Monad}},label={NonadaptiveMonadic}]
trait Monad[F[_]] {
  def bind[A, B](fa: F[A])(f: A => F[B])
  def point[A](a: A): F[A]
}
object Monad {
  implicit def futureMonad(implicit ec: ExecutionContext) = new Monad[Future] {
    def bind[A, B](fa: Future[A])(f: A => Future[B]) = fa.flatMap(f)
    def point[A](a: A): Future[A] = Future(a)
  }
}

case class Monadic[F[_], A](fa: F[A]) {
  def apply[B](handler: A => F[B])(implicit monad: Monad[F]): F[B] = {
    monad.bind(fa)(handler)
  }
}
\end{lstlisting}

\lstinline{Monadic} is an LDK more generic than \lstinline{Await}, able to ``extract'' any monadic value, not only future, as long as the corresponding \lstinline{Monad} type class instance exists.

\subsection{Collaborative library-defined keywords}\label{Collaborative library-defined keywords}

In previous sections, we ported Python's compiler-defined keywords \lstinline{yield} and \lstinline{await} to Scala, as library-defined keywords. However, those keywords are not collaborative. LDK \lstinline{Yield} and \lstinline{Await} implemented in previous sections cannot be present in the same function, while Python 3.5 allows using \lstinline{yield} and \lstinline{await} together to create asynchronous generators \cite{pep525}.

In this section, we will present a use case of Python's \lstinline{yield} and \lstinline{await} in one function, and then modify the previous implementation of LDK \lstinline{Yield} and \lstinline{Await} to gain the same ability of collaboration as Python.

\begin{lstlisting}[language=Python,style=Python3,caption={Downloading two web pages as an asynchronous generator in Python},label={download_two_pages_generator-Python}]
async def download_two_pages_generator() -> AsyncGenerator[bytes, None]:
    session = aiohttp.ClientSession()
    response1 = await session.get('http://example.com')
    content1 = await response1.read()
    yield content1
    response2 = await session.get('http://example.net')
    content2 = await response2.read()
    yield content2
\end{lstlisting}

\Cref{download_two_pages_generator-Python} shows an example of downloading two web pages with combination of \lstinline{yield} and \lstinline[language=Python,style=Python3]{await}. In Python, when an \lstinline{async} function like \lstinline{download_two_pages_generator} contains both \lstinline{yield} and \lstinline{await} keywords, the return type becomes \lstinline{AsyncGenerator}.

The corresponding type of \lstinline{AsyncGenerator[bytes, None]} in Scala could be \lstinline{Stream[Future[ByteString]]}, which should be the return type of \lstinline{apply} in the modified version of \lstinline{Yield} and \lstinline{Await}. Therefore, the modified version of \lstinline{Yield} and \lstinline{Await} can be implemented as \cref{Yield-StreamFuture,Await-StreamFuture}, and the usage of asynchronous generator with !-notation is shown in \cref{downloadTwoPages-StreamFuture}.

\begin{lstlisting}[caption={Implementing modified version of \lstinline{Yield} LDK for creating asynchronous generators},label={Yield-StreamFuture}]
case class Yield[A](element: A) {
  def apply(handler: Unit => Stream[Future[A]])(implicit ec: ExecutionContext): Stream[Future[A]] = {
    new Stream.Cons(Future(element), handler(()))
  }
}
\end{lstlisting}

\begin{lstlisting}[caption={Implementing a modified version of \lstinline{Await} LDK for creating asynchronous generators},label={Await-StreamFuture}]
case class Await[A](future: Future[A]) {
  def apply[B](handler: A => Stream[Future[B]])(implicit ec: ExecutionContext): Stream[Future[B]] = {
    val ff = future.map(handler)
    new Stream.Cons(ff.flatMap(_.head), result(ff, Duration.Inf).tail)
  }
}
\end{lstlisting}

\begin{lstlisting}[caption={Downloading two web pages as an asynchronous generator, in the style of !-notation},label={downloadTwoPages-StreamFuture}]
def downloadTwoPagesGenerator(): Stream[Future[ByteString]] = {
  // The following Await and Yield LDKs will create a Future to download the page at example.com, as the first element of the output Stream
  val response1 = !Await(Http().singleRequest(HttpRequest(HttpMethods.GET, "http://example.com")))
  val content1 = !Await(response1.entity.toStrict(timeout = 5.seconds))
  !Yield(content1.data)

  // The following Await and Yield LDKs will create a Future to download the page at example.net, as the second element of the output Stream
  val response2 = !Await(Http().singleRequest(HttpRequest(HttpMethods.GET, "http://example.net")))
  val content2 = !Await(response2.entity.toStrict(timeout = 5.seconds))
  !Yield(content2.data)

  // Remaining elements after yielded futures
  Stream.empty[Future[ByteString]]
}
\end{lstlisting}

Semantically, each \lstinline{Yield} LDK ``prepend'' a value at the head of the output \lstinline{Stream}, and the remaining parts of the output \lstinline{Stream} is a \lstinline{Stream.empty}. Any asynchronous \lstinline{Await} operations performed before a \lstinline{Yield} are collected as the asynchronous \lstinline{Future} for the yielded element.

The modified version of \lstinline{Yield} and \lstinline{Await} LDKs are collaborative, as they are both available for the domain of \lstinline{Stream[Future[ByteString]]}, thus they can be used together in one function.

\subsection{Adaptive library-defined keywords}\label{Adaptive library-defined keywords}

In previous sections, we presented two different implementations of \lstinline{Yield} in \cref{NonadaptiveYield,Yield-StreamFuture}, and two different implementations of \lstinline{Await} in \cref{NonadaptiveAwait,Await-StreamFuture}, for creating asynchronous value and asynchronous generators, respectively. However, the collaborative version of \lstinline{Yield} and \lstinline{Await} still lack of adaptivity, as the semantics the \lstinline{Yield} and \lstinline{Await} are not automatically determined by their context like Python. In this section, we will introduce the type class \lstinline{Dsl} for creating adaptive library-defined keywords to solve the adaptivity problem.

In \textit{Dsl.scala}, the \lstinline{Dsl} type class as defined in \cref{Dsl} is usually used along with \lstinline{Keyword} (\cref{Keyword}), which should be the super type of all adaptive LDKs.

\begin{lstlisting}[caption={\lstinline{Keyword}, the super type of all adaptive LDKs},label={Keyword}]

trait Keyword[Self, Value] { this: Self =>
  @inline def cpsApply[Domain](handler: Value => Domain)(implicit dsl: Dsl[Self, Domain, Value]): Domain = {
    dsl.cpsApply(this, handler)
  }

  def apply[Domain](handler: Value => Domain)(implicit dsl: Dsl[Self, Domain, Value]): Domain = cpsApply(handler)
}
\end{lstlisting}

An \lstinline{apply} call is an alias of \lstinline{cpsApply}, which registers a callback to handle the \lstinline{Value}, and finally returns a \lstinline{Domain}. The self type (\lstinline{Self}) and the value of the keyword (\lstinline{Value}) are defined in sub types of \lstinline{Keyword}. The actually implementation of a keyword is resolved by the multi-parameter type class \lstinline{Dsl}, which varies according to \lstinline{Domain}, which is the return type of the enclosing function of the keyword's call site. For example, the adaptive version of \lstinline{Yield} and \lstinline{Await} can be defined as \cref{Yield,Await}.

\begin{lstlisting}[caption={The \lstinline{Yield} LDK, the adaptive version},label={Yield}]
case class Yield[A](element: A) extends Keyword[Yield[A], Unit]
\end{lstlisting}

\begin{lstlisting}[caption={The \lstinline{Await} LDK, the adaptive version},label={Await}]
case class Await[Value](future: Future[Value]) extends Keyword[Await[Value], Value]
\end{lstlisting}

When performing !-notation on a \lstinline{Keyword} to produce a \lstinline{Value} inside a function whose return type is \lstinline{Domain}, the type class instance of \lstinline{Dsl[Keyword, Domain, Value]} is required. For example, adaptive version of LDKs in \cref{NonadaptiveYield,NonadaptiveAwait,Yield-StreamFuture,Await-StreamFuture} requires \lstinline{Dsl} instances implemented in \cref{Yield-generator-instance,Yield-async-generator-instance,Await-future-instance,Await-async-generator-instance}.

\begin{lstlisting}[caption={The \lstinline{Dsl} type class instance of \lstinline{Yield} for creating generators},label={Yield-generator-instance}]
implicit def yieldDsl[A, B >: A]: Dsl[Yield[A], Stream[B], Unit] =
  new Dsl[Yield[A], Stream[B], Unit] {
    def cpsApply(keyword: Yield[A], mapper: Unit => Stream[B]): Stream[B] = {
      new Stream.Cons(keyword.element, mapper(()))
    }
  }
\end{lstlisting}

\begin{lstlisting}[caption={The \lstinline{Dsl} type class instance of \lstinline{Yield} for creating asynchronous generators},label={Yield-async-generator-instance}]
implicit def futureYieldDsl[A, B >: A]: Dsl[Yield[A], Stream[Future[B]], Unit] =
  new Dsl[Yield[A], Stream[Future[B]], Unit] {
    def cpsApply(keyword: Yield[A], handler: Unit => Stream[Future[B]]): Stream[Future[B]] = {
      new Stream.Cons(Future.successful(keyword.element), handler(()))
    }
  }
\end{lstlisting}

\begin{lstlisting}[caption={The \lstinline{Dsl} type class instance of \lstinline{Await} for creating asynchronous values},label={Await-future-instance}]
implicit def awaitDsl[A, B](implicit ec: ExecutionContext): Dsl[Await[A], Future[B], A] =
  new Dsl[Await[A], Future[B], A] {
    def cpsApply(keyword: Await[A], handler: A => Future[B]): Future[B] = {
      keyword.future.flatMap(handler)
    }
  }
\end{lstlisting}

\begin{lstlisting}[caption={The \lstinline{Dsl} type class instance of \lstinline{Await} for creating asynchronous generators},label={Await-async-generator-instance}]
implicit def streamAwaitDsl[A, B](implicit ec: ExecutionContext): Dsl[Await[A], Stream[Future[B]], A] =
  new Dsl[Await[A], Stream[Future[B]], A] {
    def cpsApply(keyword: Await[A], handler: A => Stream[Future[B]]): Stream[Future[B]] = {
      val ff = keyword.future.map(handler)
      new Stream.Cons(ff.flatMap(_.head), result(ff, Duration.Inf).tail)
    }
  }
\end{lstlisting}

By introducing the type class \lstinline{Dsl}, the calls to \lstinline{Keyword} are ad-hoc polymorphic to the specific domain of the call site. As a result, library-defined keywords like \lstinline{Yield} and \lstinline{Await} are now adaptive like first-class keywords.

\section{Implementation}\label{Implementation}

We implemented the LDK approach in the Scala library \textit{Dsl.scala}, which consists of the following parts:

\begin{description}
  \item[The core library] contains the definition of the \lstinline{Dsl} type class and \lstinline{Keyword}, the common super type of LDKs. They are slightly different from the definition in \cref{Dsl,Keyword}:
  \begin{itemize}
    \item There is an additional dummy method \lstinline{unary_!} annotated as \lstinline{@shift} defined in \lstinline{Keyword}. The \lstinline{unary_!} method (or any other \lstinline{@shift}-annotated methods) will be specially treated by our compiler plug-ins, and it will be considered as an ordinary method for !-notation, from the point view of type checker when our compiler plug-ins are not enabled. The definition of the \lstinline{unary_!} method is especially useful for IntelliJ IDEA \footnote{https://www.jetbrains.com/idea/}'s built-in type checker, preventing the edit window in the IDE from being red marked, even though the type checker does not load compiler plug-ins.
    \item \lstinline{Keyword} is a universal trait \footnote{https://docs.scala-lang.org/overviews/core/value-classes.html}, allowing its subtypes to be value classes, which involves lower memory overhead in most of LDK use cases.
  \end{itemize}
  \item[Compiler plug-ins] performs CPS-transformation as described in \cref{CPS transformation}. There are two compiler plug-ins in \textit{Dsl.scala}: \lstinline{ResetEverywhere} and \lstinline{BangNotation}. The \lstinline{ResetEverywhere} plug-in adds a hidden \lstinline{@reset} annotation to the code block of every method in source code, and the \lstinline{BangNotation} plug-in perform CPS-transformation according to the \lstinline{unary_!} method (or any method annotated as \lstinline{@shift}) and \lstinline{@reset} annotation, which are equivalent to \lstinline{shift} and \lstinline{reset} control operators \cite{danvy1989functional}, respectively.

  In addition to block expressions mentioned in \cref{cps-block}, all other first-class control flows in Scala \footnote{Note that the \lstinline{for} expression is not converted as it is not a first-class control flow but a group of nested method calls in AST (Abstract Syntax Tree) of the Scala compiler.} are transformed to CPS form by the \lstinline{BangNotation} plug-in in the metacontinuation \cite{Danvy1990AbstractingC} approach.

  Unlike other typed delimited continuation implementations, the \lstinline{BangNotation} plug-in performs name-based CPS-transformation. Each !-notation in a transformed function can be converted to an arbitrary \lstinline{cpsApply} method call as long as it accepts a callback function parameter. Type checking for the transformed function will be performed once the transformation is done.

  Although the \lstinline{Dsl} type class does not allow changing the domain of a DSL code block, the \lstinline{BangNotation} plug-in itself allows domain changing when the \lstinline{cpsApply} method is implemented without \lstinline{Dsl} type class. Thus, the \lstinline{printf} problem can be trivially resolved by our compiler plug-ins as described in \cref{resolve-printf-problem}.

  \item[Built-in library-defined keywords] are shipped with \textit{Dsl.scala}, to provide many language features that are not available natively in Scala, including:
  \begin{itemize}
    \item The \lstinline{Await} LDK for asynchronous programming with Scala \lstinline{Future}, similar to the \lstinline[language=Python,style=Python3]{await} and \lstinline[language=Python,style=Python3]{async} keywords in C\#, Python and JavaScript.
    \item The \lstinline{Shift} LDK for asynchronous programming with delimited continuations, similar to the \lstinline{shift} operator in Scala Continuations.
    \item The \lstinline{AsynchronousIo} LDKs for perform I/O on an asynchronous channel.
    \item The \lstinline{Yield} LDK for generating lazy streams, similar to the \lstinline[language=Python,style=Python3]{yield} keyword in C\#, Python and JavaScript.
    \item The \lstinline{Each} LDK for traversing each element of a collection, similar to \lstinline{for}, \lstinline{yield} keywords for Scala collections.
    \item The \lstinline{Continue} LDK to skip an element in a LDK-based collection comprehension, similar to \lstinline{continue} keyword in many languages.
    \item The \lstinline{Fork} LDK for duplicating current thread, similar to the \lstinline{fork} system call in POSIX.
    \item The \lstinline{AutoClose} LDK to automatically close resources when exiting a scope, similar to the destructor feature in C++.
    \item The \lstinline{Monadic} LDK for creating Scalaz \cite{kenji2017scalaz} or Cats \cite{typelevel2017cats} monadic control flow, similar to the !-notation in Idris \cite{brady2013idris}.
  \end{itemize}
  \item[Asynchronous task utilities] contains a \lstinline{Task} type and related utility functions, for stack-safe asynchronous programming with the ability of exception handling and auto-closeable resource management. \lstinline{Task} is a type alias of delimited continuation whose answer type is composed of \lstinline{TailRec} and \lstinline{Throwable} in the approach described in \cref{Dsl derivation}, and the use case of our \lstinline{Task} can be found in \cref{Asynchronous programming}.

  According to the result of the benchmarks shown in \cref{Benchmarks}, the computational performance of \lstinline{Task} in \textit{Dsl.scala} is comparable to state-of-the-art Scala asynchronous programming libraries when running in HotSpot Server VM, and it achieves significant higher performance than state-of-the-art libraries when running in GraalVM.
\end{description}

\section{The underscore trick}\label{The underscore trick}

As described in \cref{Implementation}, our compiler plug-ins automatically perform \lstinline{reset} control operation for every function. However, a complex continuation is usually executed across multiple functions, which requires an approach to prevent the automatically performed \lstinline{reset} control operation.

We will propose two approaches to resolve the problem. The first solution is called the ``underscore trick'', which will be discussed in this section. Another solution is automatically derived \lstinline{Return} LDK, which will be described in \cref{Dsl derivation}.

For example, in addition to \lstinline{yield}, Python generators also allow the \lstinline{return} and \lstinline{yield from} keywords. A generator that contains both \lstinline{yield} and \lstinline{return} keywords can be invoked by \lstinline{yield from} from another generator. The elements being \lstinline{yield}ed in the former generator will be added into the latter generator, and the return value of the former generator can be used in the latter generator, too. An example of \lstinline{return} and \lstinline{yield from} is shown in \cref{returnable_generator}.

\begin{lstlisting}[language=Python,style=Python3,caption={Use \lstinline{yield from} and \lstinline{return} in Python generators},label={returnable_generator}]

def returnable_generator() -> Generator[str, None, int]:
  yield 'inside returnable_generator'
  return 1

def generator_test() -> Iterator[str]:
  yield 'before returnable_generator'
  v = yield from returnable_generator()
  yield 'after returnable_generator'
  yield f'the return value of returnable_generator is {v}'

# Output:
#   before returnable_generator
#   inside returnable_generator
#   after returnable_generator
#   the return value of returnable_generator is 1
print(*generator_test(), sep='\n')
\end{lstlisting}

Unlike generators introduced in \cref{Implementing LDKs as ordinary delimited continuations}, \lstinline{returnable_generator} has the additional ability of returning values, therefore its return type becomes \lstinline{Generator[str, None, int]}, where \lstinline{str} is the iterator element type and \lstinline{int} is the type to return \footnote{Note that the declared return type and the type to return are different in Python generators. In other words, the \lstinline{return} keyword in Python ``lifts'' the plain value to a \lstinline{Generator}.}.

When porting \lstinline{return} and \lstinline{yield from} to Scala, the return type should indicate both the element type and the type to return, thus \lstinline{Stream} is not applicable for return type any more. We can instead use the return type \lstinline{Continuation[Stream[String], Int]}, as shown in \cref{returnableGenerator}. It accepts a callback function \lstinline{k}, which can handle the \lstinline{Int} value being returned and resume the rest program in \lstinline{generatorTest}. Note that the underscore character is a Scala parameter placeholder for the callback function of the created \lstinline{Continuation} closure.

\begin{lstlisting}[caption={Returning an additional value in LDK-based generators},label={returnableGenerator}]
def returnableGenerator(): Continuation[Stream[String], Int] = _ {
  !Yield("inside returnableGenerator")
  1
}

def generatorTest(): Stream[String] = {
  !Yield("before returnableGenerator")
  val v = !Shift(returnableGenerator())
  !Yield("after returnableGenerator")
  !Yield(s"the return value of returnableGenerator is $v")
  Stream.empty
}

generatorTest.foreach(println)
\end{lstlisting}

We also create \lstinline{Shift}, an additional ad-hoc polymorphic LDK used in \lstinline{generatorTest}, to perform the continuation \footnote{There is an implicit conversion from \lstinline{Continuation} to \lstinline{Shift} LDK in \textit{Dsl.scala}, thus the explicit \lstinline{Shift()} call can be omitted. We keep the explicit instantiation of \lstinline{Shift} in this section for clarity.}. It can be considered as the LDK-based replacement of Python's \lstinline{yield from} keyword, which is defined as \cref{Shift}.

\begin{lstlisting}[caption={The definition of \lstinline{Shift} LDK},label={Shift}]
case class Shift[Domain, Value](continuation: Continuation[Domain, Value]) extends Keyword[Shift[Domain, Value], Value]
\end{lstlisting}

As described in \cref{Adaptive library-defined keywords}, we had split the LDK declaration (i.e. a subtype of \lstinline{Keyword}) from its implementation (i.e. a \lstinline{Dsl} type class instance). A \lstinline{Dsl} type class instance of \lstinline{Dsl[Shift[Stream[String], Int], Stream[String], Int]} is required to perform \lstinline{!Shift} in the domain of \lstinline{Stream[String]}. The implementation should forward \lstinline{cpsApply} call to the underlying \lstinline{continuation} of the \lstinline{Shift} LDK, as shown in \cref{shiftDsl}.

\begin{lstlisting}[caption={The \lstinline{Dsl} instance of \lstinline{Shift} LDK, to forward \lstinline{cpsApply} to the underlying \lstinline{continuation}},label={shiftDsl}]
implicit def shiftDsl[Domain, Value] =
  new Dsl[Shift[Domain, Value], Domain, Value] {
    def cpsApply(keyword: Shift[Domain, Value], handler: Value => Domain) =
      keyword.continuation(handler)
  }
\end{lstlisting}

The \lstinline{Shift} LDK can be considered as a simple wrapper of \lstinline{Continuation} that forward \lstinline{cpsApply} calls to the underlying \lstinline{continuation}.

Semantically, the automatically performed \lstinline{reset} control operator is prevented by the prepending underscore character. We call this usage of the underscore character the ``underscore trick''.

This ``underscore trick'' can also be applied on not only monomorphic delimited continuation, but also polymorphic delimited continuation \cite{asai2007polymorphic}. More examples can be found in \cref{The prefix problem}.

\section{\lstinline{Dsl} Derivation}\label{Dsl derivation}

Another solution to allow continuations to cross functions is \lstinline{Dsl} Derivation.

In \cref{Adaptive library-defined keywords}, we have present how to create an LDK for different domains, interpreted by different implementations of \lstinline{Dsl} type class instances. In this section, we will discuss derived \lstinline{Dsl} type class instances for an LDK, available in derived domains.

A derived domain means a domain whose type signature contains another domain, and a derived \lstinline{Dsl} means a \lstinline{Dsl} whose implementation internally invokes another \lstinline{Dsl}. For example, the domain \lstinline{Continuation[Stream[String], Int]}, which we used in \cref{The underscore trick}, can be considered as a derived domain of \lstinline{Stream[String]}. We will present a derived \lstinline{Dsl} to automatically ``lift'' the original domain \lstinline{Stream[String]} to the derived domain \lstinline{Continuation[Stream[String], Int]}, instead of manually creating CPS functions in the ``underscore trick''. In addition, \lstinline{Dsl} derivation approach supports early return, which is impossible in ``underscore trick''.

For example, the native keyword \lstinline{return} in Python can early return from a function, as shown in \cref{early_generator}.

\begin{lstlisting}[language=Python,style=Python3,caption={Use \lstinline{yield from} and \lstinline{return} in Python generators},label={early_generator}]
def early_generator(early_return: bool) -> Generator[str, None, int]:
  yield 'inside early_generator'
  if early_return:
    yield 'early return'
    return 1
  yield 'normal return'
  return 0

def early_generator_test() -> Iterator[str]:
  yield 'before early_generator'
  v = yield from early_generator(True)
  yield 'after early_generator'
  yield f'the return value of early_generator is {v}'

# Output:
#   before early_generator
#   inside early_generator
#   early return
#   after early_generator
#   the return value of early_generator is 1
print(*early_generator_test(), sep='\n')
\end{lstlisting}

The ability of early return is impossible with Scala native keyword \lstinline{return}, because the above \lstinline{return 0} does not compile in a function whose type is not a \lstinline{Int}. Instead we defined a new \lstinline{Return} LDK to port Python \lstinline{return} to Scala, as shown in \cref{Return,earlyGenerator}.

\begin{lstlisting}[caption={The definition of \lstinline{Return} LDK},label={Return}]
case class Return[A](returnValue: A) extends Keyword[Return[A], Nothing]
\end{lstlisting}

\begin{lstlisting}[caption={Use \lstinline{Shift} and \lstinline{Return} in LDK-based generators},label={earlyGenerator}]
def earlyGenerator(earlyReturn: Boolean): Continuation[Stream[String], Int] = {
  !Yield("inside earlyGenerator")
  if (earlyReturn) {
    !Yield("early return")
    !Return(1)
  }
  !Yield("normal return")
  !Return(0)
}

def earlyGeneratorTest(): Stream[String] = {
  !Yield("before earlyGenerator")
  val v = !Shift(earlyGenerator(true))
  !Yield("after earlyGenerator")
  !Yield(s"the return value of earlyGenerator is $v")
  Stream.empty
}

earlyGeneratorTest.foreach(println)
\end{lstlisting}

Unlike the ``underlying trick'' approach, the domain of LDKs used in \lstinline{earlyGenerator} is the return type \lstinline{Continuation[Stream[String], Int]}, which requires some \lstinline{Dsl} instances listed below:

\begin{enumerate}
  \item \lstinline{Dsl[Yield[String], Stream[String], Unit]} \\ (required by \lstinline{!Yield} in \lstinline{earlyGeneratorTest})
  \label{DslYield}

  \item \lstinline{Dsl[Shift[Stream[String], Int], Stream[String], Int]} \\ (required by \lstinline{!Shift} in \lstinline{earlyGeneratorTest})
  \label{DslShift}
  
  \item \lstinline{Dsl[Yield[String], Continuation[Stream[String], Int], Unit]} \\ (required by \lstinline{!Yield} in \lstinline{earlyGenerator})
  \label{DslYieldContinuation}
  
  \item \lstinline{Dsl[Return[Int], Continuation[Stream[String], Int], Nothing]} \\ (required by \lstinline{!Return} in \lstinline{earlyGenerator})
  \label{DslReturn}
\end{enumerate}

As discussed in \cref{The underscore trick}, \cref{DslYield,DslShift} can be resolved by \lstinline{yieldDsl} and \lstinline{shiftDsl}, respectively, but \Cref{DslYieldContinuation,DslReturn} are instances that have not been defined.
 
\Cref{DslYieldContinuation} should register a callback \lstinline{handler} and then return a new continuation, whose answer type is \lstinline{Stream[String]}. Thus, the \lstinline{Yield[String]} keyword can be performed inside the newly created continuation, interpreted by the existing \lstinline{Dsl} instance \lstinline{yieldDsl[String, String]}. The extracted value \lstinline{v} and the final handler \lstinline{k} is then passed to \lstinline{handler} to continue the execution of rest program, as shown in \cref{yieldContinuationDsl}.

\begin{lstlisting}[caption={The derived \lstinline{Dsl} instance for \lstinline{Yield} LDK, which can be used in a \lstinline{Continuation}},label={yieldContinuationDsl}]
implicit def yieldContinuationDsl = {
  new Dsl[Yield[String], Continuation[Stream[String], Int], Unit] {
    def cpsApply(keyword: Yield[String], handler: Unit => Continuation[Stream[String], Int]): Continuation[Stream[String], Int] = { k =>
      val v = !keyword
      handler(v)(k)
    }
  }
}
\end{lstlisting}

As described in \cref{Implementation}, \lstinline{!keyword} will be desugared to \lstinline[mathescape=true]|keyword.cpsApply { v => $\hdots$ }|, which is equivalent to \lstinline[mathescape=true]|yieldDsl[String, String].cpsApply(keyword, { v => $\hdots$ })| after inlining. Therefore, \lstinline{yieldContinuationDsl} can be considered as a derived \lstinline{Dsl} instance of implicitly resolved \lstinline{yieldDsl}.

Also the implementation of \lstinline{yieldContinuationDsl} can be generalized to not only \lstinline{Yield} LDK, but also any other LDKs, since \lstinline{yieldContinuationDsl} does not depend on internal details of \lstinline{Yield}. Any instances of  \lstinline{Dsl[Keyword, State => Domain, Value]} can be derived from \lstinline{Dsl[Keyword, Domain, Value]} as shown in \cref{derivedFunction1Dsl}. The implementation is the same to \cref{yieldContinuationDsl} except we manually desugared \lstinline{!keyword} and switched the instance to a more generalized type signature.

\begin{lstlisting}[caption={Derived \lstinline{Dsl} instance in a curried function},label={derivedFunction1Dsl}]
implicit def derivedFunction1Dsl[Keyword, State, Domain, Value](
  implicit restDsl: Dsl[Keyword, Domain, Value]
): Dsl[Keyword, State => Domain, Value] =
  new Dsl[Keyword, State => Domain, Value] {
    def cpsApply(keyword: Keyword, handler: Value => State => Domain): State => Domain = { k =>
      restDsl.cpsApply(keyword, { v =>
        handler(v)(k)
      })
    }
  }
\end{lstlisting}

Now the required \lstinline{Dsl} instance of \cref{DslYieldContinuation} can be resolved from either \lstinline{yieldContinuationDsl}, or, more generically, \lstinline{derivedFunction1Dsl[Yield[String], Int => Stream[String], Stream[String], Unit](yieldDsl[String, String])}. 

Similarly, since a \lstinline{!Return} LDK immediately returns from the current function, the implementation of \lstinline{Dsl} instance for \lstinline{!Return} should skip the rest part of the function, which is captured as a callback function passed to \lstinline{cpsApply}, as shown in \cref{returnDsl}. Then, \cref{DslReturn} can be resolved as \lstinline{derivedContinuationDsl(returnDsl)}.

\begin{lstlisting}[caption={The \lstinline{Dsl} instance of \lstinline{Return} LDK, to skip the registered callback function},label={returnDsl}]
implicit def returnDsl[A] =
  new Dsl[Return[A], A, Nothing] {
    def cpsApply(keyword: Return[A], handler: Nothing => A) =
      keyword.returnValue
  }
\end{lstlisting}

\lstinline{Dsl} derivation enables heterogeneous LDKs to be present in one function, whose return type is a derived domain composed from the required domain of LDKs in use. We provided more examples of this approach, including multiple mutable states, asynchronous tasks, and some advanced varieties of collection comprehension, as shown in \cref{Multiple mutable states,Asynchronous programming,Collection comprehensions}.

\section{Related works}

Previous works related to \textit{Dsl.scala} can be divided into two categories:

\begin{description}
  \item[Generic protocols of control flow operators] are motivated by the goal similar to our \lstinline{Dsl} type class. Operators of specific purposes can be implemented in single protocol, therefore, users of those operators can use a common interface for different domains. Monads and CPS functions are notable examples of such protocols.
  \item[Direct style notations] provide similar syntaxes to our name-based CPS transformation. Those notations allow users to write sequential imperative style code that will be translated to CPS or monadic style that consist of nested closures. \lstinline{yield}, \lstinline{async} / \lstinline{await}, \lstinline{reset} / \lstinline{shift}, \lstinline{for}-comprehension, \lstinline{do}-notation and !-notation are notable examples of such notations.
\end{description}

\subsection{Generators}

A generator is a special procedure to lazily produce values, which can be consumed as an iterator. Early implementation of generators are shipped in Alphard \cite{shaw1977abstraction} and CLU \cite{liskov1977abstraction}, and the feature is now available in Python, ECMAScript, C\#, and many other programming languages.

The execution of a generator will be paused at the \lstinline{yield} statement, and can be resumed when the consumer side of the generator asks for the next value. The \lstinline{yield} statement can be considered as a direct style notation for producer / consumer pattern.

Generators can be used for creating eDSLs in the following approach:

\begin{itemize}
  \item The producer side \lstinline{yield}s command objects of the DSL.
  \item The consumer side interprets these produced command objects to actually perform operations.
\end{itemize}

However, the producer side can be only executed once, therefore generators cannot represent eDSLs for collection comprehension or ``thread'' forking, though they are supported in our approach, as described in \cref{Parallel execution,Collection comprehensions}.

Another limitation of producer / consumer approach is that the type of the command object is a part of the protocol, and must be determined in advance. Therefore, the number of available commands in a generator is fixed. A generator eDSL is not composable with other generator eDSLs. In addition, generators are traditionally implemented as a first class feature by the compiler, thus they do not collaborate with other direct style notation unless changing the compiler.

In contrast, our LDK-based generators can be used along with other LDKs, including but not limited to \lstinline{Shift} (\cref{The underscore trick}), \lstinline{Return} (\cref{Dsl derivation}), \lstinline{Await} (\cref{Collaborative library-defined keywords}), \lstinline{Each} (\cref{Generator comprehensions}), without modifying the compiler or existing \lstinline{Dsl} implementations.

\subsection{\lstinline{async} and \lstinline{await}}

\lstinline{async} and \lstinline{await} are compiler-defined keywords in Python, ECMAScript, or C\#, to compose multiple asynchronous tasks into one task. Similar to generators, \lstinline{async} and \lstinline{await} provide a special purpose direct style notation, which does not support forking and does not collaborate with other direct style notations, unless modifying the implementation of the compiler like \cite{pep525} did.

Alternatively, we provide \lstinline{Await,Shift} LDK in \textit{Dsl.scala} for asynchronous programming with Scala Futures and \lstinline{Continuation}, respectively. Those asynchronous LDKs collaborate with other LDK as demonstrated in \cref{Collaborative library-defined keywords,Asynchronous programming,Asynchronous comprehensions}.

\subsection{Delimited continuations}

Delimited continuations operators of \lstinline{shift} and \lstinline{reset} \cite{Danvy1990AbstractingC} are direct style notations for performing CPS-transformation. The underlying data structures are either monomorphic \cite{danvy1989functional} or polymorphic \cite{asai2007polymorphic}, which can be considered as a generic protocol of control flow operators.

Our \lstinline{BangNotation} compiler plug-ins described in \cref{Implementation} can be considered as a simplified version of delimited continuations operators, disallowing delimited continuations across multiple functions. Fortunately, this limitation can be overcame by the ``underscore trick'' as described in \cref{The underscore trick}.

An ordinary delimited continuation is a CPS function whose implementation is predetermined. In contrast, we introduced the \lstinline{Dsl} type class as an ad-hoc polymorphic CPS function, adaptive to the enclosing domain, as described in \cref{Adaptive library-defined keywords}.

\subsection{\lstinline{for}-comprehension}\label{for-comprehension}

\lstinline{for}-comprehension is a Scala language feature, originally used to produce collections. It is a general form of list comprehension. The Scala compiler internally translates \lstinline{for}-comprehension expressions into method calls to \lstinline{map}, \lstinline{flatMap} and \lstinline{withFitler}. Like our \lstinline{BangNotation} compiler plug-in, The translation is also name-based, therefore, \lstinline{for}-comprehension can be used not only for collection generation, but also as a general direct style notation for asynchronous programming \footnote{\href{https://docs.scala-lang.org/sips/futures-promises.html}{SIP-14 - Futures and Promises}}, resource management \footnote{\href{http://jsuereth.com/scala-arm/}{Scala ARM}}, or creating monadic expression \cite{kenji2017scalaz,typelevel2017cats,twitter2016algebird}.

However, complex imperative procedures that contain native Scala control flow statements are not supported in \lstinline{for}-comprehensions. In addition, \lstinline{for}-comprehensions always end with a \lstinline{map}, preventing tail call optimization when composing multiple \lstinline{for}-comprehensions, consuming more memory and resulting worse computational performance than manually written \lstinline{flatMap} calls, according to our benchmarks in \cref{The performance impact of direct style DSLs}.

\subsection{\lstinline{do}-notation}\label{do-notation}

\lstinline{do}-notation was originally introduced in Haskell \cite{jones1998haskell} as a direct style notation for creating monadic expressions in an imperative style. \lstinline{do}-notation in Idris or \lstinline{RebindableSyntax} in Haskell are name-based, as they can be used with type classes other than monads.

The \lstinline{<-} expression is similar to the \lstinline{shift} operator in first-class delimited continuations. However, programs written in \lstinline{do}-notation can be unnecessarily verbose, since \lstinline{<-} is not an expression that can be nested in other expressions, instead, each \lstinline{<-} must be present in an individual statement.

\subsection{!-notation}

!-notation is a direct style notation in Idris \cite{brady2013idris}, to make up nested expressions in an effectful block. Our \lstinline{BangNotation} and \lstinline{ResetEverywhere} compiler plug-ins are re-implementation of Idris's !-notation in Scala, with some minor differences. Our compiler plug-ins support more native control flow expressions, including \lstinline{do}/\lstinline{while} and \lstinline{try}/\lstinline{catch}/\lstinline{finally}/\lstinline{throw}.

Since Idris's !-notation is also name-based, our \lstinline{Dsl} type class can be port to Idris and work with !-notation as well.

\subsection{Monads}\label{Monads}

A monad is a generic protocol of control flow operators used in Haskell and many other functional programming languages. A monad defines two primary operators for creating monadic expressions of a certain type.
\begin{enumerate*}
  \item The \lstinline[language=Haskell,deletekeywords={return}]{return} operator \footnote{Also called \lstinline{point} or \lstinline{pure}.} lifts a plain value to a monadic value.
  \item The \lstinline{>>=} operator \footnote{Also called \lstinline{flatMap} or \lstinline{bind}.} composes two steps monadic expressions into one monadic value, where the second step is a handler to ``flat-map'' the value of the first step into a new monadic value.
\end{enumerate*}

Since a monad is specified to a certain monadic data type, the capacity of a monadic data type is predetermined, unless introducing an additional abstract layer of interpreters. For example, the \lstinline{List} monad in Haskell \footnote{A Haskell \lstinline{List} is lazy by default, equivalent to a Scala \lstinline{Stream}.} can be used to create a \lstinline{List} based on \lstinline{List}s, but it cannot create a \lstinline{List} based on other collection types, nor creating a \lstinline{List} from a generator.

In our LDK approach, we remove the limitation of monads by separating the concept of monadic value into two orthogonal concepts: domain (concept~\ref{domain-definition}) and LDK (concept~\ref{ldk-definition}). A domain is the return type of the enclosing function, and an LDK is an operation allowed in the domain. Therefore, any collections can be created from any collections or generators, because in our approach the types of the source collection and the output collection are not necessarily the same, as demonstrated in \cref{Heterogeneous comprehensions,Generator comprehensions}.

In addition, decoupling domains and LDKs can lead simpler implementation. Our state LDKs can be used to create ordinary functions with multiple mutable states, while \lstinline{State} and \lstinline{StateT} monads are more complicated, as discussed in \cref{Single mutable state,Multiple mutable states}. Also, the implementation of a \lstinline{Cont} monad \cite{dyvbig2007monadic}, as defined in \cref{return_k,>>=_k}, is more complicated, as it creates two more additional closures -- $\lambda\kappa.t_1\left(\lambda v.t_2 v\kappa\right)$ and $\lambda v.t_2 v\kappa$ -- for each \lstinline{>>=} operator. In contrast, our \lstinline{Shift} LDK for delimited continuation runs as a simple forwarder, which creates no closure, as defined in \cref{shiftDsl} of \cref{The underscore trick}.

\begin{align}
\label{return_k}
return_k &= \lambda t.\lambda \kappa.\kappa t\\
\label{>>=_k}
\texttt{>>=}_k &= \lambda t_1.\lambda t_2.\lambda \kappa.t_1\left(\lambda v.t_2 v\kappa\right)
\end{align}

In fact, the \lstinline{>>=} operator of a monad is equivalent to a special case of \lstinline{Dsl} when the domain type and the LDK type are the same, and the \lstinline[language=Haskell,deletekeywords={return}]{return} operator of a monad can be considered another special case of \lstinline{Dsl} when the LDK is a \lstinline{Return}, which holds a plain value of the domain type, as discussed in \cref{Dsl derivation}.

There are other workarounds to overcome the limitation of monads, which will be discussed in \cref{Monad transformers,Effect handlers}.

\subsubsection{Monad transformers}\label{Monad transformers}

Monad transformers \cite{liang1995monad} are monads derived from other monads. ``Monad transformer'' is to ``monad'' as ``\lstinline{Dsl} derivation'' is to ``\lstinline{Dsl}''. The monadic data type can be composed at type level as a chain of monad transformers, so that various operations can be lifted to the same nested transformed monadic data type, which can be then used in a single monadic code block. The process to perform an operation in a monad transformer requires two steps:

\begin{enumerate}
  \item Performing the derived \lstinline{lift}, in order to transform an operation to the lifted monadic value.
  \item Performing the derived \lstinline{>>=}, in order to reduce to the final monadic value.
\end{enumerate}

However, as \cite{kiselyov2013extensible} pointed out, lifting an atomic type to a deeply nested transformed monadic data type is inefficient, because each step has to iterate through derived type class stack. The overhead of lifting a single operation is high if there is a large number of nested monad transformers.

The inefficient lifting can be avoided in our \lstinline{Dsl} derivation approach, since an LDK already represents an operation for any compatible domains. Only one step -- the \lstinline{cpsApply} method -- in the \lstinline{Dsl} type class is derived, as shown in \cref{dsl-architecture}. The LDK is executed in one step without creating an intermediate nested monadic data value. The performance improvement can be observed in the benchmark at \cref{The performance of collection manipulation in direct style DSLs}.

\subsubsection{Effect handlers}\label{Effect handlers}

Effect handler \cite{kiselyov2013extensible} is an alternative approach to monad transformers. An eDSL code block is considered as a script of \lstinline{Eff}, which is composed of effects. A generic \lstinline{Eff} monad type class instance composes individual effects into a larger script \lstinline{Eff}, which is then interpreted by a stack of \lstinline{Handler}s for each type of effect.

The effect handler approach is more efficient than monad transformer because \lstinline{Eff} is a light-weight script instead of the underlying data structure. Lifting an atomic effect to an \lstinline{Eff} is faster than lifting real data structures. However, this approach lacks of straightforwardness and keyword-wise extensibility in comparison to our LDK approach.

\begin{description}
  \item[Straightforwardness] determines how easy an eDSL is to interoperate with native types and functions of the hosting language. Effect handlers are not straightforward because \lstinline{Eff} is an additional intermediate script, which is unable to directly collaborate with the hosting language, instead, every eDSL written \lstinline{Eff} requires two steps of type classes, \lstinline{Monad} and \lstinline{Handler}, in order to produce native data structures, as shown in \cref{eff-architecture}. What is worse is, two \lstinline{Eff}s cannot invoke each other if they contain different effect stacks.

  Our LDK approach is more straightforward, as only one type class \lstinline{Dsl} is used to interpret the eDSL, as shown in \cref{dsl-architecture}. Instead of producing indirect scripts, LDKs directly produces the underlying data structures, which can be easily used in the hosting language. In addition, an eDSL block can be used in another eDSL block of a different domain as long as the underlying data types are compatible. For example, in \cref{The underscore trick}, the \lstinline{generatorTest} function can internally call the \lstinline{returnableGenerator} function even when the return types of the two functions are different.

  \item[Keyword-wise Extensibility] determines whether a new keyword or operator can be introduced into an eDSL without changing the original domain type. Unfortunately, the effect handler approach is not extensible in keyword-wise because an \lstinline{Eff} consists of a stack of effects, and each effect is specially designed for only a fix number of supported operators. It is only extensible in the domain-wise by appending a new effect, which will change the return type of an \lstinline{Eff} script.

  In contrast, our LDK approach is extensible in both domain-wise and keyword-wise.
  \begin{enumerate}
    \item Domain-wise extensibility is achieved by \lstinline{Dsl} derivation as described in \cref{Dsl derivation};
    \item Keyword-wise extensibility is achieved by providing a \lstinline{Dsl} instance for the new LDK and the existing domain. For example, we introduced \lstinline{Yield} LDK in LDK-based collection comprehension in \cref{gccFlagBuilder} without changing the return type nor the implementation of existing \lstinline{Each} LDK.
  \end{enumerate}

\end{description}

\section{Haskell implementation}\label{Haskell implementation}

We ported \textit{Dsl.scala} to Haskell as the package \textit{control-dsl} \footnote{https://hackage.haskell.org/package/control-dsl/docs/Control-Dsl.html}. The \lstinline{Dsl} type class in Haskell is defined in \cref{Dsl-haskell}. The type of \lstinline{cpsApply} is slightly different from Scala version \lstinline{Dsl} defined in \cref{Dsl}, as \lstinline{k} is an arity-2 type parameter while \lstinline{Keyword} is a first-order type parameter \footnote{
  We use shorter identifiers in \textit{control-dsl} to confirm Haskell naming conventions, as shown below:
  
  \begin{tabular}{l|l}
    Identifiers in \textit{Dsl.scala} & Identifiers in \textit{control-dsl} \\
    \hline
    \texttt{Keyword} & \texttt{k} \\
    \texttt{Domain} & \texttt{r} \\
    \texttt{Value} & \texttt{a} \\
    \texttt{Continuation} & \texttt{Cont} \\
    \texttt{handler} & \texttt{f} \\
  \end{tabular}
}. The additional type parameters for \lstinline{k} improve the ability of type inference in \lstinline{do} notation.

\begin{lstlisting}[float=htbp,language={Haskell},caption={\lstinline{Dsl} type class in \textit{control-dsl}},label={Dsl-haskell}]
class Dsl k r a where
  cpsApply :: k r a -> (a -> r) -> r
\end{lstlisting}

We also provided some helper functions for \lstinline{Dsl} based \lstinline{do}-notation, as shown in \cref{Dsl-do}. \lstinline{RebindableSyntax} language extension is required to enable those functions for \lstinline{do}-notation.

\begin{lstlisting}[float=htbp,language={Haskell},caption={Helpers for \lstinline{Dsl} based \lstinline{do}-notation},label={Dsl-do}]
(>>=) k = cpsApply k
k >> a = k >>= const a

data Return r' r a where Return :: r' -> Return r' r Void

return r = Return r >>= absurd
fail r = return (userError r)
\end{lstlisting}

Unfortunately, the additional \lstinline{r} type parameter prevents \lstinline{Dsl} derivation, since the an LDK whose type is \lstinline{k r a} can only be present in \lstinline{do} blocks of type \lstinline{r}. As a result, we are not able to port derived \lstinline{Dsl}s to Haskell like \lstinline{instance Dsl k r a => Dsl k (s -> r) a}.

To allow derived keywords, we introduced a new type class \lstinline{PolyCont}, which looses the restriction in \lstinline{Dsl}. Instead of deriving \lstinline{Dsl}, an LDK author creates derived \lstinline{PolyCont}, and finally resolves \lstinline{Dsl} from derived \lstinline{PolyCont}, as shown in \cref{PolyCont-derivation}.

\begin{lstlisting}[float=htbp,language={Haskell},caption={\lstinline{PolyCont} derivation},label={PolyCont-derivation}]
class PolyCont k r a where
  runPolyCont :: k r' a -> (a -> r) -> r

instance PolyCont k r a => PolyCont k (s -> r) a where
  runPolyCont k f s = runPolyCont k (\a -> f a s)

instance PolyCont k r a => Dsl k r a where
  cpsApply = runPolyCont
\end{lstlisting}

\lstinline{Yield} and \lstinline{Return} LDKs are ported to Haskell with the help of \lstinline{PolyCont}, as shown in \cref{PolyCont-Return-Yield-Get}

\begin{lstlisting}[float=htbp,language={Haskell},caption={\lstinline{PolyCont} instances for \lstinline{Get}, \lstinline{Yield} and \lstinline{Return}},label={PolyCont-Return-Yield-Get}]
data Get r a where Get :: forall s r. Get r s

instance PolyCont Get (s -> r) s where
  runPolyCont Get f s = f s s
  
data Yield x r a where Yield :: x -> Yield x r ()

instance PolyCont (Yield x) [x] () where
  runPolyCont (Yield x) f = x : f ()

instance PolyCont (Return r) r Void where
  runPolyCont (Return r) _ = r
\end{lstlisting}

We created another \lstinline{Dsl} instance for monomorphic delimited continuation \lstinline{Cont}, which is used to created control flow operators with nested \lstinline{do}-notation, as shown in \cref{Cont-instance,when}.

\begin{lstlisting}[float=htbp,language={Haskell},caption={\lstinline{Dsl} instance for \lstinline{Cont}},label={Cont-instance}]
newtype Cont r a = Cont { runCont :: (a -> r) -> r }

instance Dsl Cont r a where
  cpsApply = runCont
\end{lstlisting}

\begin{lstlisting}[float=htbp,language={Haskell},caption={Control flow operator \lstinline{when}},label={when}]
when :: Bool -> Cont r () -> Cont r ()
when True k = k
when False _ = Cont ($ ())
\end{lstlisting}

With the help of the above control flow operators, we are able to create direct style DSL in \lstinline{do}-notation, as shown in \cref{nested-do}

\begin{lstlisting}[float=htbp,language={Haskell},caption={Nested \lstinline{Dsl} \lstinline{do} blocks},label={nested-do}]
f = do
  Yield "foo"
  config <- Get @Bool
  when config $ do
    Yield "bar"
    return ()
  return "baz"
\end{lstlisting}

\lstinline{f} is a \lstinline{do} block that contains LDKs \lstinline{Yield}, \lstinline{Get} and \lstinline{Return} (invoked by \lstinline{return} internally). With the help of built-in \lstinline{PolyCont} instances for those keywords, \lstinline{f} can be used as a function that accepts a boolean parameter, as shown in \cref{f-pure}

\begin{lstlisting}[float=htbp,language={Haskell},caption={Running \lstinline{f} purely in REPL},label={f-pure}]
> f False :: [String]
["foo","baz"]

> f True :: [String]
["foo","bar","baz"]
\end{lstlisting}

In fact, \lstinline{f} can be any types as long as \lstinline{PolyCont} instances for the types are provided. The type can be inferred by GHC, as shown in \cref{type-f}

\begin{lstlisting}[float=htbp,language={Haskell},caption={The inferred type of a \lstinline{do} block},label={type-f}]
> :type f
f :: (PolyCont (Yield [Char]) r (),
      PolyCont (Return [Char]) r Void, PolyCont Get r Bool) =>
      r
\end{lstlisting}

For example, \lstinline{f} can be interpreted as an impure \lstinline{IO ()} (\cref{run-impure}), providing the instances defined in \cref{impure-instances}.

\begin{lstlisting}[float=htbp,language={Haskell},caption={Custom effectful instances for built-in LDKs},label={impure-instances}]
instance PolyCont (Yield String) (IO ()) () where
  runPolyCont (Yield a) = (Prelude.>>=) (putStrLn $ "Yield " ++ a)
instance PolyCont Get (IO ()) Bool where
  runPolyCont Get f = putStrLn "Get" Prelude.>> f False
instance PolyCont (Return String) (IO ()) Void where
  runPolyCont (Return r) _ = putStrLn $ "Return " ++ r
\end{lstlisting}

\begin{lstlisting}[float=htbp,language={Haskell},caption={Running \lstinline{f} effectfuly in REPL},label={run-impure}]
> f :: IO ()
Yield foo
Get
Return baz
\end{lstlisting}

In brief, the Haskell implementation \textit{control-dsl} can infer type better than \textit{Dsl.scala}, while the \lstinline{do}-notation is more verbose than !-notation in \textit{Dsl.scala}.

\section{Conclusion}

We have presented a novel approach to create direct style embedded domain specific languages that are more extensible, more straightforward and more efficient than existing monad based and continuation based approaches. The main highlights of our approaches are:

\begin{enumerate}
  \item the ability to define LDKs that work with existing native types, as if they are first-class features;
  \item the extensibility in both keyword-wise and domain-wise;
  \item \lstinline{Dsl} derivation, allowing an LDK to be adaptive to various domains.
\end{enumerate}

The capacity of LDKs is the superset of both monads and ordinary delimited continuations, thus LDKs can be used in various domains as they can be, including asynchronous or parallel programming, lazy stream generation, collection manipulation, resource management, etc. But unlike monads or ordinary delimited continuations, an LDK user can use multiple LDKs for different domains at once, along with ordinary control flow and ordinary types. No manually lifting is required, just like first-class features. This approach has been implemented in both Scala and Haskell, and can be implemented in Idris or other languages that support type classes or implicit parameters.

\subsection{Future work}

Two types of polymorphism are involved in this paper. We implemented a \lstinline{BangNotation} Scala compiler plug-in to perform name-based CPS-transformation, which support answer type modification, or \textbf{polymorphic delimited continuation}; we introduced \lstinline{Dsl} type class, which allows running an LDK as a CPS function adaptive to the predetermined answer type, or \textbf{ad-hoc polymorphic delimited continuation}. In the future, we will investigate how to represent a delimited continuation that is both polymorphic and ad-hoc polymorphic.

\appendix

\section{Use cases}

We will present some use cases of name-based CPS transformation and LDK in this section, to illustrate the simplicity of our approach, in comparison to previous solutions.

\subsection{Resolve the \lstinline{printf} problem, trivially}\label{resolve-printf-problem}

The type-safe \lstinline{printf} problem \cite{danvy1998functional} is often used to demonstrate the ability of modifying the answer type of a typed delimited continuation. The problem can be also resolved by \textit{Dsl.scala}'s CPS-transformation plug-ins as shown in \cref{printf}.

\begin{lstlisting}[caption={A solution of the type-safe \lstinline{printf} problem in \textit{Dsl.scala}},label={printf}]
object IntPlaceholder {
  @shift def unary_! : String = ???
  def cpsApply[Domain](f: String => Domain): Int => Domain = { i: Int =>
    f(i.toString)
  }
}

object StringPlaceholder {
  @shift def unary_! : String = ???
  def cpsApply[Domain](f: String => Domain): String => Domain = f
}

def f1 = "Hello World!"
def f2 = "Hello " + !StringPlaceholder + "!"
def f3 = "The value of " + !StringPlaceholder + " is " + !IntPlaceholder + "."

println(f1) // Output: Hello World!
println(f2("World")) // Output: Hello World!
println(f3("x")(3)) // Output: The value of x is 3.
\end{lstlisting}

This solution works because our plug-ins performs CPS-transformation for \lstinline{f1}, \lstinline{f2} and \lstinline{f3}, as shown in \cref{transformed-printf}.

\begin{lstlisting}[caption={The translated source code of \lstinline{Dsl.scala}-base solution of \lstinline{printf} problem},label={transformed-printf}]
// The type of f1 is inferred as `String`
def f1 = "Hello World!"

// The type of f2 is inferred as `String => String`
def f2 = StringPlaceholder.cpsApply { tmp =>
  "Hello " + tmp + "!"
}

// The type of f3 is inferred as `String => Int => String`
def f3 = StringPlaceholder.cpsApply { tmp0 =>
  IntPlaceholder.cpsApply { tmp1 =>
    "The value of " + tmp0 + " is " + tmp1 + "."
  }
}
\end{lstlisting}

Our solution is more concise than the solution with Scala Continuations \cite{rompf2009implementing}, because:
\begin{enumerate*}
  \item No explicit \lstinline{reset} is required, as \lstinline{reset} is automatically added by the \lstinline{ResetEverywhere} plug-in.
  \item No explicit \lstinline{@cps} type annotation is required, since the \lstinline{BangNotation} plug-in is name-based. The type of \lstinline{f1}, \lstinline{f2} and \lstinline{f3} can be inferred automatically, according to Scala's type inference algorithm for closures.
\end{enumerate*}

\subsection{The prefix problem}\label{The prefix problem}

The \lstinline{prefix} problem introduced in \cite{asai2007polymorphic} is a problem that requires polymorphic delimited continuations, which is a CPS function whose answer type can be modified, i.e. \lstinline{(A => B) => C} where \lstinline{B} and \lstinline{C} differ. We provide a \lstinline{PolymorphicShift} LDK to perform \lstinline{shift} control operator for polymorphic delimited continuations, as defined in \cref{PolymorphicShift}. Note that \lstinline{PolymorphicShift} is not interpreted by \lstinline{Dsl}, hence it is not ad-hoc polymorphic.

\begin{lstlisting}[caption={The definition of PolymorphicShift},label={PolymorphicShift}]
final case class PolymorphicShift[A, B, C](cpsApply: (A => B) => C) {
  @shift def unary_! : A = ???
}

implicit def implicitPolymorphicShift[A, B, C](cpsApply: (A => B) => C) = PolymorphicShift(cpsApply)  
\end{lstlisting}

The solution to \lstinline{prefix} problem uses ``underscore trick'' along with \lstinline{PolymorphicShift}, as shown in \cref{prefix}.

\begin{lstlisting}[caption={The solution to \lstinline{prefix} problem by the ``underscore trick''},label={prefix}]
def visit[A](lst: List[A]): (List[A] => List[A] @reset) => List[List[A]] = _ {
  lst match {
    case Nil =>
      !{ (h: List[A] => List[A]) =>
        Nil
      }
    case a :: rest =>
      a :: !{ (k: List[A] => List[A] @reset) =>
        k(Nil) :: k(!visit(rest))
      }
  }
}

def prefix[A](lst: List[A]) = !visit(lst)

// Output: List(List(1), List(1, 2), List(1, 2, 3))
println(prefix(List(1, 2, 3)))
\end{lstlisting}

Traditional polymorphic delimited continuations performs CPS transformation across functions. In contrast, with the help of the ``underscore trick'', we achieve the same ability of answer type modification as polymorphic delimited continuations, by performing function-local CPS-translation.

\subsection{Mutable states}

Purely functional programming languages usually do not support first-class mutable variables. In those languages, mutable states can be implemented in state monads. In this section, we will present an alternative approach based on LDK to simulate mutable variable in a pure language \footnote{Scala is an impure language, but we don't use Scala's native \lstinline{var} or other impure features when simulating mutable states, therefore, our approach can be ported to Haskell or other pure languages as described in \cref{Haskell implementation}.}. Unlike state monads, our LDK-based approach is more straightforward, and supports multiple mutable states without manually lifting.

\subsubsection{Single mutable state}\label{Single mutable state}

We use unary function as the domain of mutable state. The parameter of the unary function can be read from \lstinline{Get} LDK, and changed by \lstinline{Put} LDK, which are defined in \cref{Get,Put}, respectively.

\begin{lstlisting}[caption={The definition of \lstinline{Get} LDK},label={Get}]
case class Get[S]() extends Keyword[Get[S], S]
\end{lstlisting}

\begin{lstlisting}[caption={The definition of \lstinline{Put} LDK},label={Put}]
case class Put[S](value: S) extends Keyword[Put[S], Unit]
\end{lstlisting}

\Cref{upperCasedLastCharacter} is an example of a unary function that accepts a string parameter and returns the upper-cased last character of the parameter. The initial value is read from \lstinline{Get} LDK, then it is changed to upper-case by \lstinline{Put} LDK. At last, another \lstinline{Get} LDK is performed to read the changed value, whose last character is then returned.

\begin{lstlisting}[caption={Using \lstinline{Get} and \lstinline{Put} in a unary function},label={upperCasedLastCharacter}]
def upperCasedLastCharacter: String => Char = {
  val initialValue = !Get[String]()
  !Put(initialValue.toUpperCase)

  val upperCased = !Get[String]()
  Function.const(upperCased.last)
}

// Output: O
println(upperCasedLastCharacter("foo"))
\end{lstlisting}

The \lstinline{Dsl} instances for \lstinline{Get} and \lstinline{Put} used in \lstinline{upperCasedLastCharacter} are shown in \cref{getDsl,putDsl}. The \lstinline{Dsl} instance for \lstinline{Get} LDK passes the \lstinline{currentValue} to the \lstinline{handler} of current LDK, and then continues the enclosing unary function; the \lstinline{Dsl} instance for \lstinline{Put} LDK ignores \lstinline{previousValue} and continues the enclosing unary function with the new \lstinline{value} in \lstinline{Put}.

\begin{lstlisting}[caption={The \lstinline{Dsl} instance for \lstinline{Get} LDK},label={getDsl}]

implicit def getDsl[S0, S <: S0, A] =
  new Dsl[Get[S0], S => A, S0] {
    def cpsApply(keyword: Get[S0], handler: S0 => S => A): S => A = { currentValue =>
      handler(currentValue)(currentValue)
    }
  }
\end{lstlisting}

\begin{lstlisting}[caption={The \lstinline{Dsl} instance for \lstinline{Put} LDK},label={putDsl}]
implicit def putDsl[S0, S >: S0, A] =
  new Dsl[Put[S0], S => A, Unit] {
    def cpsApply(keyword: Put[S0], handler: Unit => S => A): S => A = { previousValue =>
      handler(())(keyword.value)
    }
  }
\end{lstlisting}

Traditionally, the data type of state monad is an opaque type alias of \lstinline{S => (S, A)}, which is more complicated than our domain type \lstinline{S => A}, indicating state monads are potentially less efficient than LDK-based implementation. We will discuss the reason why monad-based DSL are more complicated and less efficient than LDK-based DSL in \cref{Monads}.

\subsubsection{Multiple mutable states}\label{Multiple mutable states}

\lstinline{Get} and \lstinline{Put} LDKs can be performed on multiple mutable states as well. The domain types are curried functions in those use cases.

In \cref{formatter}, we present an example to create a \lstinline{formatter} that performs \lstinline{Put} on a \lstinline{Vector[Any]} to store parts of the string content. At last, a \lstinline{Return} LDK is performed at last to concatenate those parts. The \lstinline{formatter} internally performs \lstinline{Get} LDKs of different types to retrieve different parameters.

\begin{lstlisting}[caption={Using \lstinline{Get} and \lstinline{Put} in a curried function},label={formatter}]
def formatter: Double => Int => Vector[Any] => String = {
  !Put(!Get[Vector[Any]] :+ "x=")
  !Put(!Get[Vector[Any]] :+ !Get[Double])
  !Put(!Get[Vector[Any]] :+ ",y=")
  !Put(!Get[Vector[Any]] :+ !Get[Int])

  !Return((!Get[Vector[Any]]).mkString)
}

// Output: x=0.5,y=42
println(formatter(0.5)(42)(Vector.empty))
\end{lstlisting}

Since we had introduced \lstinline{Dsl} instance for \lstinline{Get} and \lstinline{Put} LDKs in unary functions, now we only need a derived \lstinline{Dsl} instance to port these LDKs in curried functions, which is defined in \cref{derivedFunction1Dsl}.

By combining \lstinline{getDsl} and \lstinline{derivedFunction1Dsl} together, the Scala compiler automatically searches matched type in the curried function when resolving the implicit \lstinline{Dsl} instance for a \lstinline{Get} LDK. For example, \lstinline{!Get[Vector[Any]]()} reads the third parameter of the \lstinline{formatter}. It will be translated to \lstinline[mathescape=true]|Get[Vector[Any]]().cpsApply { _ => $\hdots$ }|, where the \lstinline{cpsApply} call requires an instance of type \lstinline{Dsl[Get[Vector[Any]], Double => Int => Vector[Any] => String, Vector[Any]]}, which will be resolved as \lstinline{derivedFunction1Dsl(derivedFunction1Dsl(getDsl))}. Similarly, the \lstinline{Dsl} instance for reading the first parameter and the second parameter can be resolved as \lstinline{getDsl} and \lstinline{derivedFunction1Dsl(getDsl)}, respectively.

Derived \lstinline{Dsl} instance for \lstinline{Put} and \lstinline{Return} can be resolved similarly. Since all the \lstinline{!Put} LDK in \lstinline{formatter} write the third parameter, their \lstinline{Dsl} instances are \lstinline{derivedFunction1Dsl(derivedFunction1Dsl(putDsl))}; the \lstinline{Dsl} instance for \lstinline{!Return} are \lstinline{derivedFunction1Dsl(derivedFunction1Dsl(derivedFunction1Dsl(returnDsl)))}.

Now we had demonstrated a simple and straightforward solution for the feature of multiple mutable states, with the help of nested \lstinline{Dsl} derivation.

\subsection{Asynchronous programming}\label{Asynchronous programming}

With the help of \lstinline{Dsl} derivation, a complex DSL can be composed of simple features. In this section we will present a sophisticated implementation of asynchronous task, which is composed of three independent features:
\begin{enumerate*}
  \item asynchronous result handling
  \item exception handling
  \item stack safety
\end{enumerate*}, as defined in \cref{Task}. A new infix type alias \lstinline{!!} is used instead of \lstinline{Continuation}, as a shorter notation for nested \lstinline{Continuation} types.

\begin{lstlisting}[caption={The definition of asynchronous \lstinline{Task}},label={Task}]
type !![Domain, Value] = Continuation[Domain, Value]
type Task[A] = TailRec[Unit] !! Throwable !! A
\end{lstlisting}

\lstinline{Task} supports the features of tail call optimization and exception handling. Each feature corresponds to a part the type signature. \lstinline{scala.util.control.TailCalls.TailRec} is used for tail call optimization, and \lstinline{scala.Throwable} is used to represent the internal exceptional state.

We create some derived \lstinline{Dsl}s to handle exceptions, which support domains whose types match the pattern of \lstinline[mathescape=true]{($L_i$ !! $\hdots$ !! $L_0$ !! Throwable !! $R_0$ !! $\hdots$ !! $R_i$)}, and some derived \lstinline{Dsl}s to optimize tail calls as trampolines, which support domains whose types match the pattern of \lstinline[mathescape=true]{(TailRec[$\hdots$] !! $R_0$ !! $\hdots$ !! $R_i$)}, where $L_0 \hdots L_i$ and $R_0 \hdots R_i$ are arbitrary number of types \footnote{Those \lstinline{Dsl}s are implemented in the \lstinline{Dsl} derivation technique described in \cref{Dsl derivation}. Check the artifact for the complete implementation}. Therefore, \lstinline{Dsl} instances for \lstinline{Task} are composed from these orthogonal features.

In \cref{An asynchronous HTTP client}, we will present how to create an asynchronous HTTP client from \lstinline{Task}; in \cref{Parallel execution}, we will introduce the usage of \lstinline{Task[Seq[A]]}, which collects the results of multiple tasks into a \lstinline{Seq}, either executed in parallel or sequentially.

\subsubsection{An asynchronous HTTP client}\label{An asynchronous HTTP client}

\Cref{httpClient} is an example of an HTTP client built from low-level Java NIO.2 asynchronous IO operations. Note that the ``underscore trick'' is used to allow \lstinline{Task} to be executed across functions.

\begin{lstlisting}[caption={An asynchronous HTTP client},label={httpClient}]
def readAll(channel: AsynchronousByteChannel, destination: ByteBuffer): Task[Unit] = _ {
  if (destination.remaining > 0) {
    val numberOfBytesRead: Int = !Read(channel, destination)
    numberOfBytesRead match {
      case -1 =>
      case _  => !readAll(channel, destination)
    }
  } else {
    throw new IOException("The response is too big to read.")
  }
}

def writeAll[Domain](channel: AsynchronousByteChannel, destination: ByteBuffer): Task[Unit] = _ {
  while (destination.remaining > 0) {
    !Write(channel, destination)
  }
}

def asynchronousHttpClient(url: URL): Task[String] = _ {
  val socket = AsynchronousSocketChannel.open()
  try {
    val port = if (url.getPort == -1) 80 else url.getPort
    val address = new InetSocketAddress(url.getHost, port)
    !Connect(socket, address)
    val request = ByteBuffer.wrap(s"GET ${url.getPath} HTTP/1.1\r\nHost:${url.getHost}\r\nConnection:Close\r\n\r\n".getBytes)
    !writeAll(socket, request)
    val MaxBufferSize = 100000
    val response = ByteBuffer.allocate(MaxBufferSize)
    !readAll(socket, response)
    response.flip()
    io.Codec.UTF8.decoder.decode(response).toString
  } finally {
    socket.close()
  }
}
\end{lstlisting}

We defined \lstinline{Connect}, \lstinline{Read} and \lstinline{Write} LDKs to register handlers to Java NIO.2 asynchronous IO operators. In addition to \lstinline{Task} domain, those LDKs also support any domains that match types of \lstinline[mathescape=true]{($\hdots$ !! Unit !! Throwable !! $\hdots$)} or \lstinline[mathescape=true]{($\hdots$ !! TailRec[Unit] !! Throwable !! $\hdots$)} \footnote{Check the artifact for complete implementation.}.

\lstinline[mathescape=true]{!readAll($\hdots$)} and \lstinline[mathescape=true]{!writeAll($\hdots$)} are equivalent to \lstinline[mathescape=true]{!Shift(readAll($\hdots$))} and \lstinline[mathescape=true]{!Shift(writeAll($\hdots$))}. The explicit \lstinline{Shift} calls are omitted because we provided an implicit conversion from any \lstinline{Continuation}s(including \lstinline{Task}s) to \lstinline{Shift} LDKs.

We also provided a \lstinline{blockingAwait} method, to block the current thread until the result of the asynchronous task is ready, therefore, \lstinline{asynchronousHttpClient} can be used synchronously, as shown in \cref{usingAsynchonousHttpClient}.

\begin{lstlisting}[caption={Using the example HTTP client},label={usingAsynchonousHttpClient}]
val httpResponse = Task.blockingAwait(asynchronousHttpClient(new URL("http://example.com/")))
httpResponse should startWith("HTTP/1.1 200 OK")
\end{lstlisting}

\subsubsection{Parallel execution}\label{Parallel execution}

Another useful LDK for asynchronous programming is \lstinline{Fork}, which duplicate the current control flow, and the child control flow are executed in parallel, similar to the POSIX \lstinline{fork} system call, as shown in \cref{usingHttpClientInParallel}.

\begin{lstlisting}[caption={Using HTTP client in parallel},label={usingHttpClientInParallel}]
val Urls = Seq(
  new URL("http://example.com/"),
  new URL("http://example.org/")
)
def parallelTask: Task[Seq[String]] = {
  val url: URL = !Fork(Urls)
  val content: String = !httpClient(url)
  !Return(content)
}

val Seq(fileContent0, fileContent1) = Task.blockingAwait(parallelTask)
assert(fileContent0.startsWith("HTTP/1.1 200 OK"))
assert(fileContent1.startsWith("HTTP/1.1 200 OK"))
\end{lstlisting}

Since the execution of \lstinline{parallelTask} is forked, the two URLs will be downloaded in parallel. The results are then collected into a \lstinline{Task} of \lstinline{Seq} at the \lstinline{!Return} LDK.

\subsubsection{Modularity and performance}

Our approach achieved both better modularity and better performance than previous implementation.

Most of previous asynchronous programing libraries, including Scala Future \cite{haller2012sip}, Monix \cite{nedelcu2017monix}, and Cats effects \cite{typelevel2017cats}, are built from a solid implementation, along with some callback scheduler for custom behaviors. In contrast, our approach separate atomic features into independent \lstinline{Dsl} type classes.

Scalaz Concurrent \cite{kenji2017scalaz} or other monad transformer \cite{liang1995monad} based approach can separate asynchronous programing features into monad of asynchronous handling and monad transformer of exception handling. Even though, trampolines are not able to implemented as monad transformers, as a result, intrusive code for trampolines must be present in each monad instance and monad transformer instance, or the call stack may overflow. In contrast, our approach allows non-intrusive \lstinline{Dsl} derivation for \lstinline{TailRec[Unit]}, then the ability of stack safety will be added on previously stack unsafe \lstinline{Dsl}s.

According to our benchmarks in \cref{Benchmarks}, on HotSpot JVM, our \lstinline{Task} implementation is much faster than monad transformer based implementations, and has similar performance in comparison to solid implementations. On GraalVM, our \lstinline{Task} implementation is faster than all other implementations.
\subsection{Collection comprehensions}\label{Collection comprehensions}

List comprehension or array comprehension is a feature to create a collection based on some other collections, which has been implemented as first class feature in many programming languages including Scala. In this section, we will present the \lstinline{Each} LDK, which allows collection comprehensions for arbitrary collection types. Unlike other first class comprehension, our LDK-based collection comprehension collaborates with other LDKs, thus allowing creating complex code of effects or actions along with collection comprehensions.

\subsubsection{Heterogeneous comprehensions}\label{Heterogeneous comprehensions}

Suppose we want to calculate all composite numbers below $n$, the program can be written in Scala's native \lstinline{for}-comprehension as shown in \cref{compositeNumbersBelow-for}.

\begin{lstlisting}[caption={Calculating all composite numbers below $n$ with \lstinline{for}-comprehension},label={compositeNumbersBelow-for}]
def compositeNumbersBelow(n: Int) = (for {
  i <- 2 until math.ceil(math.sqrt(n)).toInt
  j <- 2 * i until n by i
} yield j).to[Set]
\end{lstlisting}

The \lstinline{compositeNumbersBelow} can be ported to LDK-based collection comprehension with the following steps:

\begin{enumerate}
  \item Replacing the \lstinline{for} keyword and the trailing \lstinline[mathescape=true]{.to[$CollectionType$]} by the heading $CollectionType$.
  \item Replacing every \lstinline[mathescape=true]{$p$ <- $e$} by \lstinline[mathescape=true]{val $p$ = !Each($e$)}.
  \item Moving the value to \lstinline{yield} to the last expression position of the comprehension block.
\end{enumerate}

Therefore, \cref{compositeNumbersBelow-for} can be rewrite to \cref{compositeNumbersBelow} with the help of the \lstinline{Each} LDK, or \cref{compositeNumbersBelow-simplified} after removing the temporary variable \lstinline{j}.

\begin{lstlisting}[caption={Calculating all composite numbers below $n$ with \lstinline{Each} LDK},label={compositeNumbersBelow}]
def compositeNumbersBelow(n: Int): Set[Int] = Set {
  val i = !Each(2 until math.ceil(math.sqrt(n)).toInt)
  val j = !Each(2 * i until n by i)
  j
}

// Output: Set(10, 14, 6, 9, 12, 8, 4)
println(compositeNumbersBelow(15))
\end{lstlisting}

\begin{lstlisting}[caption={Calculating all composite numbers below $n$ with \lstinline{Each} LDK, the simplicied version},label={compositeNumbersBelow-simplified}]
def compositeNumbersBelow(n: Int): Set[Int] = Set {
  val i = !Each(2 until math.ceil(math.sqrt(n)).toInt)
  !Each(2 * i until n by i)
}
\end{lstlisting}

Note that \lstinline{compositeNumbersBelow} creates a \lstinline{Set}, which is different from the type of source collection. Our LDK-base collection comprehension allows heterogeneous source collection types. Even other collection-like types, including \lstinline{Array} and \lstinline{String}, are supported, as shown in \cref{heterogeneous}.

\begin{lstlisting}[caption={LDK-based heterogeneous collection comprehension based on \lstinline{Array} and \lstinline{String}},label={heterogeneous}]
def heterogeneous = List { !Each(Array("foo", "bar", "baz")) + !Each("LDK") }

// Output: List(fooL, fooD, fooK, barL, barD, barK, bazL, bazD, bazK)
println(heterogeneous)
\end{lstlisting}
\subsubsection{Filters}

We also provides the \lstinline{Continue} LDK to skip an element from the source collections. It provides the similar feature to the \lstinline{if} clause in Scala's native \lstinline{for}-comprehension. An example of using \lstinline{Continue} LDK to calculate prime numbers is shown in \cref{primeNumbersBelow}.

\begin{lstlisting}[caption={Calculating all prime numbers below $n$ with \lstinline{Each} and \lstinline{Continue} LDK},label={primeNumbersBelow}]
def primeNumbersBelow(maxNumber: Int) = List {
  val compositeNumbers = compositeNumbersBelow(maxNumber)
  val i = !Each(2 until maxNumber)
  if (compositeNumbers(i)) !Continue
  i
}

// Output: List(2, 3, 5, 7, 11, 13)
println(primeNumbersBelow(15))
\end{lstlisting}

The implementation of \lstinline{Continue} LDK is similar to \lstinline{Return}, except is pass an empty collection to the handler instead of the given value.

\subsubsection{Asynchronous comprehensions}\label{Asynchronous comprehensions}

The \lstinline{Each} LDK can be used in \lstinline{Task} of collections as well, with the help of \lstinline{Dsl} derivation. The usage of \lstinline{Each} is very similar to the \lstinline{Fork} keyword. The only difference is that \lstinline{Each} sequentially executes tasks while \lstinline{Fork} executes tasks in parallel. For example, if we replace the \lstinline{Fork} LDK in \cref{usingHttpClientInParallel} by \lstinline{Each}, those URLs will be fetched sequentially, as shown in \cref{usingHttpClientSequentially}.

\begin{lstlisting}[caption={Using HTTP client in parallel},label={usingHttpClientSequentially}]
def sequentialTask: Task[Seq[String]] = {
  val url: URL = !Each(Urls)
  val content: String = !httpClient(url)
  !Return(content)
}
\end{lstlisting}

\subsubsection{Generator comprehensions}\label{Generator comprehensions}

Since the \lstinline{Each} LDK works in any function that returns a collection, it can be also used in \lstinline{Stream} functions, which support the \lstinline{Yield} LDK as well. As a result, generator and collection comprehension can be used together.

Suppose we are creating a function to prepare flags for invoking the \texttt{gcc} command line tool. Given a source file and a list of include paths, it should return a \lstinline{Stream} of the command line.
It can be implemented from the \lstinline{Yield}, \lstinline{Each} and \lstinline{Continue} as shown in \cref{gccFlagBuilder}.

\begin{lstlisting}[caption={Build a command-line by using generator and collection comprehension together},label={gccFlagBuilder}]
def gccFlagBuilder(sourceFile: String, includes: String*): Stream[String] = {
  !Yield("gcc")
  !Yield("-c")
  !Yield(sourceFile)
  val include = !Each(includes)
  !Yield("-I")
  !Yield(include)
  !Continue
}

// Output: List(gcc, -c, main.c, -I, lib1/include, -I, lib2/include)
println(gccFlagBuilder("main.c", "lib1/include", "lib2/include").toList)
\end{lstlisting}

\section{Benchmarks}\label{Benchmarks}

We created some benchmarks to evaluate the computational performance of code generated by our compiler plug-in for LDKs, especially, we are interesting how our name-based CPS transformation and other direct style DSL affect the performance in an effect system that support both asynchronous and synchronous effects.

Our benchmarks measured the performance of LDKs in the \lstinline{Task} domain mentioned in \cref{Asynchronous programming}, along with other combination of effect system with direct style DSL, listed in \cref{combination}:

\begin{table}[htbp]
  \begin{tabular}{l|l}
    Effect System & direct style DSL \\
    \hline
    The \texttt{Task} LDK & name-based CPS transformation provided by \textit{Dsl.scala} \\
    Scala Future \cite{haller2012sip} & Scala Async \cite{haller2013sip} \\
    Scala Continuation library \cite{rompf2009implementing} & Scala Continuation compiler plug-in \\
    Monix tasks \cite{nedelcu2017monix} & \texttt{for}-comprehension \\
    Cats effects \cite{typelevel2017cats} & \texttt{for}-comprehension \\
    Scalaz Concurrent \cite{kenji2017scalaz} & \texttt{for}-comprehension \\
  \end{tabular}
  \caption{The combination of effect system and direct style DSL being benchmarked}
  \label{combination}
\end{table}

\subsection{The performance of recursive functions in effect systems}

The purpose of the first benchmark is to determine the performance of recursive functions in various effect system, especially when a direct style DSL is used.

\subsubsection{The performance baseline}

In order to measure the performance impact due to direct style DSLs, we have to measure the performance baseline of different effect systems at first. We created some benchmarks for the most efficient implementation of a sum function in each effect system. These benchmarks perform the following computation:

\begin{itemize}
  \item Creating a \lstinline{List[X[Int]]} of 1000 tasks, where \lstinline{X} is the data type of task in the effect system.
  \item Performing recursive right-associated ``binds'' on each element to add the \lstinline{Int} to an accumulator, and finally produce a \lstinline{X[Int]} as a task of the sum result.
  \item Running the task and blocking awaiting the result.
\end{itemize}

Note that the tasks in the list is executed in the current thread or in a thread pool. We keep each task returning a simple pure value, because we want to measure the overhead of effect systems, not the task itself.

The ``bind'' operation means the primitive operation of each effect system. For Monix tasks, Cats effects, Scalaz Concurrent and Scala Continuations, the ``bind'' operation is \lstinline{flatMap}; for \textit{Dsl.scala}, the ``bind'' operation is \lstinline{Shift} LDK, which may or may not be equivalent to \lstinline{flatMap} according to the type of the current domain. In \lstinline{Continuation} domain, the \lstinline{Dsl} instance for \lstinline{Shift} LDK is resolved as \lstinline{derivedContinuationDsl(shiftDsl)}, whose \lstinline{cpsApply} method flat maps a \lstinline{Continuation} to another \lstinline{Continuation}; when using ``underscore trick'', the \lstinline{Dsl} instance for \lstinline{Shift} LDK is resolved as \lstinline{shiftDsl}, which just forwards \lstinline{cpsApply} to the underlying CPS function as a plain function call.

We use the !-notation to perform the \lstinline{cpsApply} in \textit{Dsl.scala}. The !-notation results the exact same Java bytecode to manually passing a callback function to \lstinline{cpsApply}, as shown in \cref{RawSum.dsl}.

\begin{lstlisting}[float=htbp,caption={The most efficient implementation of sum based on ordinary CPS function},label={RawSum.dsl}]
def loop(tasks: List[Task[Int]], accumulator: Int = 0)(callback: Int => TaskDomain): TaskDomain = {
  tasks match {
    case head :: tail =>
      // Expand to: Shift(head).cpsApply(i => loop(tail, i + accumulator)(callback))
      loop(tail, !head + accumulator)(callback)
    case Nil =>
      callback(accumulator)
  }
}
\end{lstlisting}

However, direct style DSLs for other effect systems are not used in favor of raw \lstinline{flatMap} calls, in case of decay of the performance. \Cref{RawSum.future} shows the benchmark code for Scala Futures. The code for all the other effect systems are similar to it.

\begin{lstlisting}[float=htbp,caption={The most efficient implementation of sum based on Scala Futures},label={RawSum.future}]
def loop(tasks: List[Future[Int]], accumulator: Int = 0): Future[Int] = {
  tasks match {
    case head :: tail =>
      head.flatMap { i =>
        loop(tail, i + accumulator)
      }
    case Nil =>
      Future.successful(accumulator)
  }
}
\end{lstlisting}

The benchmark result is shown in \cref{RawSum} (larger score is better):

\begin{table}[htbp]
  \begin{tabular}{l|l|l|rl}
   \multicolumn{1}{c|}{\texttt{Benchmark}} & \texttt{executedIn} & \texttt{size} & \multicolumn{2}{c}{\texttt{Score, ops/s}} \\
  \hline
  \texttt{RawSum.cats} & \texttt{thread-pool} & \texttt{1000} & \texttt{799.072} & \scriptsize $\pm$ \texttt{3.094}  \\
  \texttt{RawSum.cats} & \texttt{current-thread} & \texttt{1000} & \texttt{26932.907} & \scriptsize $\pm$ \texttt{845.715}  \\
  \texttt{RawSum.dsl} & \texttt{thread-pool} & \texttt{1000} & \texttt{729.947} & \scriptsize $\pm$ \texttt{4.359}  \\
  \texttt{RawSum.dsl} & \texttt{current-thread} & \texttt{1000} & \texttt{31161.171} & \scriptsize $\pm$ \texttt{589.935}  \\
  \texttt{RawSum.future} & \texttt{thread-pool} & \texttt{1000} & \texttt{575.403} & \scriptsize $\pm$ \texttt{3.567}  \\
  \texttt{RawSum.future} & \texttt{current-thread} & \texttt{1000} & \texttt{876.377} & \scriptsize $\pm$ \texttt{8.525}  \\
  \texttt{RawSum.monix} & \texttt{thread-pool} & \texttt{1000} & \texttt{743.340} & \scriptsize $\pm$ \texttt{11.314}  \\
  \texttt{RawSum.monix} & \texttt{current-thread} & \texttt{1000} & \texttt{55421.452} & \scriptsize $\pm$ \texttt{251.530}  \\
  \texttt{RawSum.scalaContinuation} & \texttt{thread-pool} & \texttt{1000} & \texttt{808.671} & \scriptsize $\pm$ \texttt{3.917}  \\
  \texttt{RawSum.scalaContinuation} & \texttt{current-thread} & \texttt{1000} & \texttt{17391.684} & \scriptsize $\pm$ \texttt{385.138}  \\
  \texttt{RawSum.scalaz} & \texttt{thread-pool} & \texttt{1000} & \texttt{722.743} & \scriptsize $\pm$ \texttt{11.234}  \\
  \texttt{RawSum.scalaz} & \texttt{current-thread} & \texttt{1000} & \texttt{15895.606} & \scriptsize $\pm$ \texttt{235.992}  \\
  \end{tabular}
  \caption{The benchmark result of sum for performance baseline}
  \label{RawSum}
\end{table}

The \lstinline{Task} alias of continuation-passing style function used with \textit{Dsl.scala} is quite fast. \textit{Dsl.scala}, Monix and Cats Effects score on top 3 positions for either tasks running in the current thread or in a thread pool.

\subsubsection{The performance impact of direct style DSLs}\label{The performance impact of direct style DSLs}

In this section, we will present the performance impact when different syntax notations are introduced. For ordinary CPS functions, we added one more !-notation to avoid manually passing the \lstinline{callback} in the previous benchmark (\cref{LeftAssociatedSum.dsl,RightAssociatedSum.dsl}). For other effect systems, we refactored the previous sum benchmarks to use Scala Async, Scala Continuation's \lstinline{@cps} annotations, and \lstinline{for}-comprehension, respectively (\cref{LeftAssociatedSum.future,RightAssociatedSum.future,LeftAssociatedSum.scalaContinuation,RightAssociatedSum.scalaContinuation,LeftAssociatedSum.scalaz,RightAssociatedSum.scalaz}).

\begin{lstlisting}[float=htbp,caption={Left-associated sum based on LDKs of \textit{Dsl.scala}},label={LeftAssociatedSum.dsl}]
def loop(tasks: List[Task[Int]]): Task[Int] = _ {
  tasks match {
    case head :: tail =>
      !head + !loop(tail)
    case Nil =>
      0
  }
}
\end{lstlisting}

\begin{lstlisting}[float=htbp,caption={Right-associated sum based on LDKs of \textit{Dsl.scala}},label={RightAssociatedSum.dsl}]
def loop(tasks: List[Task[Int]], accumulator: Int = 0): Task[Int] = _ {
  tasks match {
    case head :: tail =>
      !loop(tail, !head + accumulator)
    case Nil =>
      accumulator
  }
}
\end{lstlisting}

\begin{lstlisting}[float=htbp,caption={Left-associated sum based on Scala Async},label={LeftAssociatedSum.future}]
def loop(tasks: List[Future[Int]]): Future[Int] = async {
  tasks match {
    case head :: tail =>
      await(head) + await(loop(tail))
    case Nil =>
      0
  }
}
\end{lstlisting}

\begin{lstlisting}[float=htbp,caption={Right-associated sum based on Scala Async},label={RightAssociatedSum.future}]
def loop(tasks: List[Future[Int]], accumulator: Int = 0): Future[Int] = async {
  tasks match {
    case head :: tail =>
      await(loop(tail, await(head) + accumulator))
    case Nil =>
      accumulator
  }
}
\end{lstlisting}

\begin{lstlisting}[float=htbp,caption={Left-associated sum based on Scala Continuation plug-in},label={LeftAssociatedSum.scalaContinuation}]
def loop(tasks: List[() => Int @suspendable]): Int @suspendable = {
  tasks match {
    case head :: tail =>
      head() + loop(tail)
    case Nil =>
      0
  }
}
\end{lstlisting}

\begin{lstlisting}[float=htbp,caption={Right-associated sum based on Scala Continuation plug-in},label={RightAssociatedSum.scalaContinuation}]
def loop(tasks: List[() => Int @suspendable], accumulator: Int = 0): Int @suspendable = {
  tasks match {
    case head :: tail =>
      loop(tail, head() + accumulator)
    case Nil =>
      accumulator
  }
}
\end{lstlisting}

\begin{lstlisting}[float=htbp,caption={Left-associated sum based on \lstinline{for}-comprehension},label={LeftAssociatedSum.scalaz}]
def loop(tasks: List[Task[Int]]): Task[Int] = {
  tasks match {
    case head :: tail =>
      for {
        i <- head
        accumulator <- loop(tail)
      } yield i + accumulator
    case Nil =>
      Task(0)
  }
}
\end{lstlisting}

\begin{lstlisting}[float=htbp,caption={Right-associated sum based on \lstinline{for}-comprehension},label={RightAssociatedSum.scalaz}]
def loop(tasks: List[Task[Int]], accumulator: Int = 0): Task[Int] = {
  tasks match {
    case head :: tail =>
      for {
        i <- head
        r <- loop(tail, i + accumulator)
      } yield r
    case Nil =>
      Task.now(accumulator)
  }
}
\end{lstlisting}

Note that reduced sum can be implemented in either left-associated recursion or right-associated recursion. The above code contains benchmark for both cases. The benchmark result is shown in \cref{LeftAssociatedSum,RightAssociatedSum}:

\begin{table}[htbp]
  \begin{tabular}{l|l|l|rl}
   \multicolumn{1}{c|}{\texttt{Benchmark}} & \texttt{executedIn} & \texttt{size} & \multicolumn{2}{c}{\texttt{Score, ops/s}} \\
  \hline
  \texttt{LeftAssociatedSum.cats} & \texttt{thread-pool} & \texttt{1000} & \texttt{707.940} & \scriptsize $\pm$ \texttt{10.497}  \\
  \texttt{LeftAssociatedSum.cats} & \texttt{current-thread} & \texttt{1000} & \texttt{16165.442} & \scriptsize $\pm$ \texttt{298.072}  \\
  \texttt{LeftAssociatedSum.dsl} & \texttt{thread-pool} & \texttt{1000} & \texttt{729.122} & \scriptsize $\pm$ \texttt{7.492}  \\
  \texttt{LeftAssociatedSum.dsl} & \texttt{current-thread} & \texttt{1000} & \texttt{19856.493} & \scriptsize $\pm$ \texttt{386.225}  \\
  \texttt{LeftAssociatedSum.future} & \texttt{thread-pool} & \texttt{1000} & \texttt{339.415} & \scriptsize $\pm$ \texttt{1.486}  \\
  \texttt{LeftAssociatedSum.future} & \texttt{current-thread} & \texttt{1000} & \texttt{410.785} & \scriptsize $\pm$ \texttt{1.535}  \\
  \texttt{LeftAssociatedSum.monix} & \texttt{thread-pool} & \texttt{1000} & \texttt{742.836} & \scriptsize $\pm$ \texttt{9.904}  \\
  \texttt{LeftAssociatedSum.monix} & \texttt{current-thread} & \texttt{1000} & \texttt{19976.847} & \scriptsize $\pm$ \texttt{84.222}  \\
  \texttt{LeftAssociatedSum.scalaContinuation} & \texttt{thread-pool} & \texttt{1000} & \texttt{657.721} & \scriptsize $\pm$ \texttt{9.453}  \\
  \texttt{LeftAssociatedSum.scalaContinuation} & \texttt{current-thread} & \texttt{1000} & \texttt{15103.883} & \scriptsize $\pm$ \texttt{255.780}  \\
  \texttt{LeftAssociatedSum.scalaz} & \texttt{thread-pool} & \texttt{1000} & \texttt{670.725} & \scriptsize $\pm$ \texttt{8.957}  \\
  \texttt{LeftAssociatedSum.scalaz} & \texttt{current-thread} & \texttt{1000} & \texttt{5113.980} & \scriptsize $\pm$ \texttt{110.272}  \\
  \end{tabular}
  \caption{The benchmark result of left-associated sum in direct style DSLs}
  \label{LeftAssociatedSum}
\end{table}

\begin{table}[htbp]
  \begin{tabular}{l|l|l|rl}
   \multicolumn{1}{c|}{\texttt{Benchmark}} & \texttt{executedIn} & \texttt{size} & \multicolumn{2}{c}{\texttt{Score, ops/s}} \\
  \hline
    \texttt{RightAssociatedSum.cats} & \texttt{thread-pool} & \texttt{1000} & \texttt{708.441} & \scriptsize $\pm$ \texttt{9.201}  \\
    \texttt{RightAssociatedSum.cats} & \texttt{current-thread} & \texttt{1000} & \texttt{15971.331} & \scriptsize $\pm$ \texttt{315.063}  \\
    \texttt{RightAssociatedSum.dsl} & \texttt{thread-pool} & \texttt{1000} & \texttt{758.152} & \scriptsize $\pm$ \texttt{4.600}  \\
    \texttt{RightAssociatedSum.dsl} & \texttt{current-thread} & \texttt{1000} & \texttt{22393.280} & \scriptsize $\pm$ \texttt{677.752}  \\
    \texttt{RightAssociatedSum.future} & \texttt{thread-pool} & \texttt{1000} & \texttt{338.471} & \scriptsize $\pm$ \texttt{2.188}  \\
    \texttt{RightAssociatedSum.future} & \texttt{current-thread} & \texttt{1000} & \texttt{405.866} & \scriptsize $\pm$ \texttt{2.843}  \\
    \texttt{RightAssociatedSum.monix} & \texttt{thread-pool} & \texttt{1000} & \texttt{736.533} & \scriptsize $\pm$ \texttt{10.856}  \\
    \texttt{RightAssociatedSum.monix} & \texttt{current-thread} & \texttt{1000} & \texttt{21687.351} & \scriptsize $\pm$ \texttt{107.249}  \\
    \texttt{RightAssociatedSum.scalaContinuation} & \texttt{thread-pool} & \texttt{1000} & \texttt{654.749} & \scriptsize $\pm$ \texttt{7.983}  \\
    \texttt{RightAssociatedSum.scalaContinuation} & \texttt{current-thread} & \texttt{1000} & \texttt{12080.619} & \scriptsize $\pm$ \texttt{274.878}  \\
    \texttt{RightAssociatedSum.scalaz} & \texttt{thread-pool} & \texttt{1000} & \texttt{676.180} & \scriptsize $\pm$ \texttt{7.705}  \\
    \texttt{RightAssociatedSum.scalaz} & \texttt{current-thread} & \texttt{1000} & \texttt{7911.779} & \scriptsize $\pm$ \texttt{79.296}  \\
  \end{tabular}
  \caption{The benchmark result of right-associated sum in direct style DSLs}
  \label{RightAssociatedSum}
\end{table}

The result demonstrates that the name-based CPS transformation provided by \textit{Dsl.scala} is faster than all other direct style DSLs in the right-associated sum benchmark. The \textit{Dsl.scala} version sum consumes a constant number of memory during the loop, because we implemented a tail-call detection in our CPS-transform compiler plug-in, and the \lstinline{Dsl} interpreter for \lstinline{Task} use a trampoline technique \cite{tarditi1992no}. On the other hand, the benchmark result of Monix Tasks, Cats Effects and Scalaz Concurrent posed a significant performance decay, because they costs O(n) memory due to the \lstinline{map} call generated by \lstinline{for}-comprehension, although those effect systems also built in trampolines. In general, the performance of recursive monadic binds in a \lstinline{for}-comprehension is always underoptimized due to the inefficient \lstinline{map}.

\subsection{The performance of collection manipulation in effect systems}

The previous sum benchmarks measured the performance of manually written loops, but usually we may want to use higher-ordered functions to manipulate collections. We want to know how those higher-ordered functions can be expressed in direct style DSLs, and how would the performance be affected by direct style DSLs.

In this section, we will present the benchmark result for computing the Cartesian product of lists.

\subsubsection{The performance baseline}

As we did in sum benchmarks, we created some benchmarks to maximize the performance for Cartesian product. Our benchmarks create the Cartesian product from \lstinline{traverseM} for Scala Future, Cats Effect, Scalaz Concurrent and Monix Tasks. \Cref{RawCartesianPruduct.future} shows the benchmark code for Scala Future.

\begin{lstlisting}[float=htbp,caption={Cartesian product for Scala Future, based on Scalaz's \lstinline{traverseM}},label={RawCartesianPruduct.future}]
def cellTask(taskX: Future[Int], taskY: Future[Int]): Future[List[Int]] = async {
  List(await(taskX), await(taskY))
}

def listTask(rows: List[Future[Int]], columns: List[Future[Int]]): Future[List[Int]] = {
  rows.traverseM { taskX =>
    columns.traverseM { taskY =>
      cellTask(taskX, taskY)
    }
  }
}
\end{lstlisting}

Scala Async or \lstinline{for}-comprehension is used in element-wise task \lstinline{cellTask}, but the collection manipulation \lstinline{listTask} is kept as manually written higher order function calls, because neither Scala Async nor \lstinline{for}-comprehension supports \lstinline{traverseM}.

The benchmark for \textit{Dsl.scala} is entirely written in LDKs as shown in \cref{RawCartesianPruduct.dsl}:

\begin{lstlisting}[float=htbp,caption={Cartesian product for ordinary CPS functions, based on \textit{Dsl.scala}},label={RawCartesianPruduct.dsl}]
def cellTask(taskX: Task[Int], taskY: Task[Int]): Task[List[Int]] = _ {
  List(!taskX, !taskY)
}

def listTask(rows: List[Task[Int]], columns: List[Task[Int]]): Task[List[Int]] = {
  cellTask(!Each(rows), !Each(columns))
}
\end{lstlisting}

The \lstinline{Each} LDK is available here because it is adaptive. \lstinline{Each} LDK can be used in not only \lstinline{List[_]} domain, but also \lstinline{(_ !! Coll[_])} domain as long as \lstinline{Coll} is a Scala collection type that supports \lstinline{CanBuildFrom} type class.

We didn't benchmark Scala Continuation here because all higher ordered functions for \lstinline{List} do not work with Scala Continuation.

The benchmark result is shown in \cref{RawCartesianProduct}.

\begin{table}[htbp]
  \begin{tabular}{l|l|l|rl}
   \multicolumn{1}{c|}{\texttt{Benchmark}} & \texttt{executedIn} & \texttt{size} & \multicolumn{2}{c}{\texttt{Score, ops/s}} \\
  \hline
  \texttt{RawCartesianProduct.cats} & \texttt{thread-pool} & \texttt{50} & \texttt{136.415} & \scriptsize $\pm$ \texttt{1.939}  \\
  \texttt{RawCartesianProduct.cats} & \texttt{current-thread} & \texttt{50} & \texttt{1346.874} & \scriptsize $\pm$ \texttt{7.475}  \\
  \texttt{RawCartesianProduct.dsl} & \texttt{thread-pool} & \texttt{50} & \texttt{140.098} & \scriptsize $\pm$ \texttt{2.062}  \\
  \texttt{RawCartesianProduct.dsl} & \texttt{current-thread} & \texttt{50} & \texttt{1580.876} & \scriptsize $\pm$ \texttt{27.513}  \\
  \texttt{RawCartesianProduct.future} & \texttt{thread-pool} & \texttt{50} & \texttt{100.340} & \scriptsize $\pm$ \texttt{1.894}  \\
  \texttt{RawCartesianProduct.future} & \texttt{current-thread} & \texttt{50} & \texttt{93.678} & \scriptsize $\pm$ \texttt{1.829}  \\
  \texttt{RawCartesianProduct.monix} & \texttt{thread-pool} & \texttt{50} & \texttt{142.071} & \scriptsize $\pm$ \texttt{1.299}  \\
  \texttt{RawCartesianProduct.monix} & \texttt{current-thread} & \texttt{50} & \texttt{1750.869} & \scriptsize $\pm$ \texttt{18.365}  \\
  \texttt{RawCartesianProduct.scalaz} & \texttt{thread-pool} & \texttt{50} & \texttt{78.588} & \scriptsize $\pm$ \texttt{0.623}  \\
  \texttt{RawCartesianProduct.scalaz} & \texttt{current-thread} & \texttt{50} & \texttt{357.357} & \scriptsize $\pm$ \texttt{2.102}  \\
  \end{tabular}
  \caption{The benchmark result of Cartesian product for performance baseline}
  \label{RawCartesianProduct}
\end{table}

Monix tasks, Cats Effects and ordinary CPS functions created from \textit{Dsl.scala} are still the top 3 scored effect systems.

\subsubsection{The performance of collection manipulation in direct style DSLs}\label{The performance of collection manipulation in direct style DSLs}

We then refactored the benchmarks to direct style DSLs. \Cref{CartesianProduct.future} is the code for Scala Future, written in \lstinline{ListT} monad transformer provided by Scalaz. The benchmarks for Monix tasks, Scalaz Concurrent are also rewritten in the similar style.

\begin{lstlisting}[float=htbp,caption={Cartesian product for Scala Future, based on \lstinline{ListT} transformer},label={CartesianProduct.future}]
def listTask(rows: List[Future[Int]], columns: List[Future[Int]]): Future[List[Int]] = {
  for {
    taskX <- ListT(Future.successful(rows))
    taskY <- ListT(Future.successful(columns))
    x <- taskX.liftM[ListT]
    y <- taskY.liftM[ListT]
    r <- ListT(Future.successful(List(x, y)))
  } yield r
}.run
\end{lstlisting}

With the help of \lstinline{ListT} monad transformer, we are able to merge \lstinline{cellTask} and \lstinline{listTask} into one function in a direct style \lstinline{for}-comprehension, avoiding any manual written callback functions.

We also merged \lstinline{cellTask} and \lstinline{listTask} in the \textit{Dsl.scala} version of benchmark as shown in \cref{CartesianProduct.dsl}.

\begin{lstlisting}[float=htbp,caption={Cartesian product for ordinary CPS functions, in one function},label={CartesianProduct.dsl}]
def listTask: Task[List[Int]] = reset {
  List(!(!Each(inputDslTasks)), !(!Each(inputDslTasks)))
}
\end{lstlisting}

This time, Cats Effects are not benchmarked due to lack of \lstinline{ListT} in Cats. The benchmark result are shown in \cref{CartesianProduct}.

\begin{table}[htbp]
  \begin{tabular}{l|l|l|rl}
   \multicolumn{1}{c|}{\texttt{Benchmark}} & \texttt{executedIn} & \texttt{size} & \multicolumn{2}{c}{\texttt{Score, ops/s}} \\
  \hline
  \texttt{CartesianProduct.dsl} & \texttt{thread-pool} & \texttt{50} & \texttt{283.450} & \scriptsize $\pm$ \texttt{3.042}  \\
  \texttt{CartesianProduct.dsl} & \texttt{current-thread} & \texttt{50} & \texttt{1884.514} & \scriptsize $\pm$ \texttt{47.792}  \\
  \texttt{CartesianProduct.future} & \texttt{thread-pool} & \texttt{50} & \texttt{91.233} & \scriptsize $\pm$ \texttt{1.333}  \\
  \texttt{CartesianProduct.future} & \texttt{current-thread} & \texttt{50} & \texttt{150.234} & \scriptsize $\pm$ \texttt{20.396}  \\
  \texttt{CartesianProduct.monix} & \texttt{thread-pool} & \texttt{50} & \texttt{28.597} & \scriptsize $\pm$ \texttt{0.265}  \\
  \texttt{CartesianProduct.monix} & \texttt{current-thread} & \texttt{50} & \texttt{120.068} & \scriptsize $\pm$ \texttt{17.676}  \\
  \texttt{CartesianProduct.scalaz} & \texttt{thread-pool} & \texttt{50} & \texttt{31.110} & \scriptsize $\pm$ \texttt{0.662}  \\
  \texttt{CartesianProduct.scalaz} & \texttt{current-thread} & \texttt{50} & \texttt{87.404} & \scriptsize $\pm$ \texttt{1.734}  \\
  \end{tabular}
  \caption{The benchmark result of Cartesian product in direct style DSLs}
  \label{CartesianProduct}
\end{table}

Despite the trivial manual lift calls in \lstinline{for}-comprehension, the monad transformer approach causes terrible computational performance in comparison to manually called \lstinline{traverseM}. In contrast, the performance of \textit{Dsl.scala} even got improved when \lstinline{cellTask} is inlined into \lstinline{listTask}.

\begin{acks}

We are very grateful to Marisa Kirisame for many helpful comments and discussions.

\end{acks}

\bibliography{bibliography}

\end{document}